\newif\ifpretty
\prettytrue
\ifpretty
\documentclass[aps,twocolumn,showpacs,preprintnumbers,floatfix,amsmath,amssymb]{revtex4}
\else
\documentclass[aps,preprint,showpacs,preprintnumbers,amsmath,amssymb]{revtex4}
\fi

\usepackage{graphicx}
\DeclareGraphicsExtensions{.eps}
\graphicspath{{./FIG/}}

\ifpretty\else

\fi

\begin{document}

\title{Temperature Dependence of the 
Hydrophobic Hydration and Interaction of 
Simple Solutes: 
An Examination of Five Popular Water Models}

\author{Dietmar Paschek}
\email[]{dietmar.paschek@udo.edu}
\homepage[]{http://ganter.chemie.uni-dortmund.de/~pas/}
\affiliation{Department of Physical Chemistry, Otto-Hahn Str. 6,
  University of Dortmund, D-44221 Dortmund, Germany}


\date{\today}

\begin{abstract}
We examine five different popular rigid water models 
(SPC, SPCE, TIP3P, TIP4P and TIP5P) 
using molecular dynamics simulations
in order to investigate 
the hydrophobic hydration 
and interaction
of apolar Lennard-Jones solutes as a
function of temperature in the range between 
$275\;\mbox{K}$ and $375\;\mbox{K}$ along the $0.1\;\mbox{MPa}$ isobar.
For all investigated models and state points
we calculate the excess chemical potential for the noble
gases and Methane 
employing the Widom particle insertion technique.
All water models
exhibit
 too small hydration entropies, but show a clear hierarchy.
TIP3P shows poorest agreement with experiment
whereas TIP5P is closest to the experimental data at lower temperatures
and SPCE is closest at higher temperatures. As a first
approximation, this behaviour can
be rationalised as a temperature-shift with respect to
the solvation behaviour found in real water.
A rescaling procedure inspired by information theory model of 
Hummer et al. ({\em Chem.Phys. }258, 349-370 (2000)) 
suggests that the different solubility curves for the
different models and real water can be largely explained on
the basis of the
different density curves at constant pressure.
In addition, the models that give a good representation of the water
structure at ambient conditions (TIP5P, SPCE and TIP4P) 
show considerably better agreement with the experimental data
than the ones which exhibit less structured O-O correlation 
functions (SPC and TIP3P).
In the second part of the paper we calculate the hydrophobic
interaction between Xenon particles directly
from a series of 60 ns simulation runs. 
We find that the temperature dependence of the association
is to a large extent related
to the strength of the solvation entropy.
Nevertheless, differences between the models seem to 
require a more detailed molecular picture.
The TIP5P model shows by far the strongest temperature dependence.
The suggested density-rescaling is also applied to the chemical
potential in the Xenon-Xenon contact-pair configuration, indicating
the presence of a temperature where the hydrophobic interaction
turns into purely repulsive.
The predicted association for Xenon in real water suggest
the presence a strong variation with temperature, comparable to the behaviour
found for TIP5P water. 
Comparing different water models and experimental data we
conclude that a proper description of density effects is
an  important requirement for a water model to account correctly
for the correct description of the hydrophobic effects.
A water model exhibiting a density maximum at
the correct temperature is desirable.
\end{abstract}

%%% Local Variables:
%%% mode: latex
%%% TeX-master: "paper"
%%% End:

\maketitle

\section{INTRODUCTION}

Nonpolar solutes show a strong tendency to aggregate 
when dissolved in water due to the relatively strong water-water
interaction in comparison to the weak solute-water interaction
\cite{Ben-Naim:Hydrophobic,Kauzmann:59,Ben-Naim:79}.
However, in addition to such energetical considerations
the hydration entropy of small simple solutes is found to
be negative, which is usually explained as increased ordering of the
molecules in the hydration shell \cite{Privalov:88,Haymet:96}. This 
is the characteristic feature of the
so called {\em hydrophobic hydration} of small apolar
particles \cite{Pierotti:76,Pratt:77}.
As a consequence, with increasing temperature
the association of hydrophobic particles is found to be enhanced 
in order to minimise the solvation entropy 
penalty \cite{Smith.D:92,Smith.D:93}.
This entropy-driven association process is usually referred to as
{\em hydrophobic interaction} and has been subject of numerous
simulation studies
\cite{Geiger:79,Zichi:85,Smith.D:92,Smith.D:93,Pearlman:93,%
Belle:93,Dang:94,Forsman:94,Luedemann:96,Skipper:96,Young:97,%
Hummer:96:1,Luedemann:97,Rick:97,Hummer:2001,%
Shimizu:2000,Rick:2000,Shimizu:2001,%
Ghosh:2001,Ghosh:2002}.

Hydrophobic effects are considered to play an important role
concerning protein stability and other self
assembly phenomena \cite{Tanford,Privalov:88,Huang:2000}. 
The strength of hydrophobic interactions, however, might 
as well determine largely the temperature-range where
the protein remains physiologically functional 
before thermal unfolding takes place.
In addition, the weakening of hydrophobic interactions at lower 
temperatures is presumably of importance for the understanding of the
cold denaturation of proteins \cite{Privalov:1990,Tsai:2002,Kumar:2003}.
This is probably of particular
importance for 
systems that show an entropy driven configurational ``folding'' transition
such as the ``hydrophobic'' collapse of polymeric elastin \cite{Urry:93,Darwin:2001.1} 
and small tropoelastin-like oligo-peptides \cite{Reiersen:98} 
around $300\,\mbox{K}-330\,\mbox{K}$.
The present status of conceptual understanding of
hydrophobic hydration and hydrophobic
interaction has been reviewed recently by
Pratt \cite{PrattRev:2002} and Southall et al. \cite{Southall:2002}, 
whereas Smith and Haymet \cite{Smith:2003} give a tutorial
overview over the currently available computational methods.

Recent methodological improvements in simulation techniques, 
such as the ``parallel tempering'' approach
(or ``replica exchange molecular dynamics'') 
give rise to the possibility of 
explicitly studying the thermal folding/unfolding
transition of 
{\em solvated} 
proteins \cite{Gnanakaran:2003,Sanbonmatsu:2002}.
Molecular dynamics simulations of proteins in aqueous solution, 
however, still largely 
depend on the use of simple model potentials for water.
Therefore it might be interesting to compare different
popular rigid water models 
with respect to hydrophobic interactions
and to show how their strengths  might be related to 
the behaviour of other known thermodynamical 
and structural properties of the model liquids. For this purpose we employ the 
three center point charge models proposed by the Berendsen group 
(SPC \cite{Berendsen:81} and SPCE \cite{Berendsen:87})
as well as three candidates representing the TIP-family of
water models according to Jorgensens group: 
The three center TIP3P and four center 
TIP4P \cite{Jorgensen:83}, as
well as the recently proposed five center TIP5P model \cite{Mahoney:2000}.

The most simple and as well most often studied hydrophobic 
model system in
this context is, of course,
{\em hydrophobic} Lennard-Jones 
particles dissolved in water \cite{Pratt:2002}.
The outline of the paper is therefore the following. 

First we would like
to examine the the performance of the five different
models with 
respect to the hydrophobic hydration behaviour as a function
of temperature with a focus on the physiologically important
temperature range between $275\,\mbox{K}$ and $375\,\mbox{K}$.
This is done a in the spirit of the
paper by Guillot and Guissani \cite{Guillot:93}
where we calculate the chemical potentials and
solvation entropies for Lennard Jones particles
representing the noble gases and Methane
and compare them with experimental data. Simulation runs
of $20\,\mbox{ns}$ allow an accurate determination of the excess chemical potential
using the Widom particle insertion method.

In the second part of the paper we study the association behaviour
of hydrophobic particles (only Xenon)
as a function of temperature for all five 
water models by conducting long ($60\,\mbox{ns}$) 
MD simulation runs of solutions of the hydrophobic
particles in water, as suggested recently by the work of Ghosh et al. 
\cite{Ghosh:2001,Ghosh:2002}.
The potential of mean force (PMF) between
the hydrophobic particles can then be obtained directly from the
pair distribution functions. Thus we are able to calculate
the hydrophobic interaction for the different models as a function
of temperature.

A very recent study on lattice models by Widom et al. \cite{Widom:2003}
suggests a linear relationship
between the strength of the hydrophobic hydration and and the 
hydrophobic pair interaction. Therefore it
might also be interesting to provide accurate data for both
on a set of realistic water models.

%%% Local Variables:
%%% mode: latex
%%% TeX-master: "paper"
%%% End:

\section{METHODS}

\subsection{MD Simulation details}

\label{sec:MD}

\ifpretty\begin{table*}[!ht]
  \centering
  \renewcommand{\arraystretch}{1.1}
  \ifpretty
  \renewcommand{\tabcolsep}{1.58cm}
  \else
  \renewcommand{\tabcolsep}{0.58cm}
  \fi
  \small
  \begin{tabular}{lccc} \\ \hline\hline \\[-6pt]
  Model &
  $\sigma/\mbox{\AA}$ &   
  $\epsilon\,k_B^{-1}/\mbox{K}$ &
  $q/\mbox{e}$ 
\\[6pt] \hline \\[-6pt]
SPC            &  $3.1656$(O)   & $78.2$(O)  & $0.41$(H) \\
SPCE           &  $3.1656$(O)   & $78.2$(O)  & $0.4238$(H) \\
TIP3P          &  $3.1506$(O)   & $76.58$(O) & $0.417$(H) \\
TIP4P          &  $3.1536$(O)   & $78.08$(O) & $0.52$(H) \\
TIP5P          &  $3.12$(O)     & $80.56$(O) & $0.241$(H)  \\
Ne             &  $3.035$  & $18.6$  & $0$ \\
Ar             &  $3.415$  & $125.0$  & $0$ \\
Kr             &  $3.675$  & $169.0$  & $0$ \\
Xe             &  $3.975$  & $214.7$  & $0$ \\
$\mbox{CH}_4$  &  $3.730$  & $147.5$  & $0$ %
\\[6pt] \hline\hline
  \end{tabular}
  \caption{\footnotesize
    Lennard-Jones potential parameters and partial charges
    describing the water-water and solute-solute pair interactions.
    The solute-water cross parameters are deduced from the
    Lorentz-Berthelot mixing rules:
    $\sigma_{ij}\!=\!\left(\sigma_{ii}+\sigma_{jj}\right)/2$,
    $\epsilon_{ij}\!=\!\sqrt{\epsilon_{ii}\epsilon_{jj}}$.
    For further information
    on the geometry of the water models we refer to
    the original
    references \cite{Berendsen:81,Berendsen:87,Jorgensen:83,Mahoney:2000}.
    The solute-solute-parameters were taken from Refs. \cite{Hirschfelder:54,Guillot:93}.}
  \label{tab:models}
\end{table*}
 \fi
We employ molecular dynamics (MD) simulations in the NPT ensemble using
the Nos\'e-Hoover thermostat 
\cite{Nose:84,Hoover:85}
and the Rahman-Parinello barostat 
\cite{Parrinello:81,Nose:83} with
coupling times $\tau_T\!=\!1.5\,\mbox{ps}$ and 
$\tau_p\!=\!2.5\,\mbox{ps}$
(assuming the isothermal compressibility to be 
$\chi_T\!=\!4.5\;10^{-5}\,\mbox{bar}^{-1}$), respectively.
The electrostatic interactions are treated
in the ``full potential'' approach
by the smooth particle mesh Ewald summation 
\cite{Essmann:95} with a real space
cutoff of $0.9\,\mbox{nm}$ and a mesh spacing of approximately
$0.12\,\mbox{nm}$ and 4th order
interpolation. The Ewald convergence factor $\alpha$ was set to
$3.38\,\mbox{nm}^{-1}$ (corresponding to a relative accuracy of
the Ewald sum of $10^{-5}$).
A $2.0\,\mbox{fs}$ 
timestep was used for all simulations and the constraints were solved
using the SETTLE procedure \cite{Miyamoto:92}.
All simulations reported here were carried out using 
the GROMACS 3.1  program \cite{gmxpaper,gmx31}.
Statistical errors in the analysis
were computed using the method of Flyvbjerg and Petersen \cite{Flyvbjerg:89}.
For all reported systems and different
statepoints initial equilibration runs of $1\,\mbox{ns}$ length 
were performed using the Berendsen 
weak coupling scheme for pressure and temperature control
$\tau_T\!=\!\tau_p\!=\!0.5\,\mbox{ps}$ \cite{Berendsen:84}. 

In order to determine the excess chemical potential of
the hydrophobic particles we performed a series of simulations generally
using 256 water molecules for all five different water models
SPC, SPCE, TIP3P, TIP4P and TIP5P (model parameters
are given in Table \ref{tab:models}).
However, the excess chemical potential is known to be
sensitive to finite size effects \cite{Siepmann:92}. In order to estimate this influence, we
additionally carried out 
simulations containing  500 and 864 water molecules but only
for the SPCE model.
All model systems were simulated at five different temperatures
$275\,\mbox{K}$, $300\,\mbox{K}$, $325\,\mbox{K}$, $350\,\mbox{K}$
and $375\,\mbox{K}$ at a pressure of $0.1\,\mbox{MPa}$. 
Each of these simulations extended to $20\,\mbox{ns}$  and $2\times 10^4$ configurations
were stored for further analysis. 
To determine the hydrophobic interaction between Xenon particles
(for the model parameters see Table  \ref{tab:models})
we use MD simulations containing 500 water molecules and
8 Xenon particles employing the same simulation parameters outlined
above. Again, the  five different water model systems are
studied at $275\,\mbox{K}$, $300\,\mbox{K}$, $325\,\mbox{K}$, $350\,\mbox{K}$
and $375\,\mbox{K}$ at a pressure of $0.1\,\mbox{MPa}$.
Here, runs over $60\,\mbox{ns}$ were conducted, while storing $7.5\times 10^4$ 
configurations for further analysis.
The simulations protocols 
showing the obtained densities and average potential energies
are  given in Table \ref{tab:sim}. 
\ifpretty\begin{table*}[!ht]
  \centering
  \ifpretty
  \renewcommand{\tabcolsep}{0.54cm}
  \renewcommand{\arraystretch}{1.0}
  \else
  \renewcommand{\tabcolsep}{0.1cm}
  \renewcommand{\arraystretch}{0.86}
  \fi
  \small
  \begin{tabular}{lccc|cc} \\ \hline\hline \\[-6pt]
  Model &
  $T/\mbox{K}$ &
  $\left< \rho\right>/\mbox{kg}\,\mbox{m}^{-3}$ &
  $\left< E\right>/\mbox{kJ}\,\mbox{mol}^{-1}$  &
  $\left< \rho\right>/\mbox{kg}\,\mbox{m}^{-3}$ &
  $\left< E\right>/\mbox{kJ}\,\mbox{mol}^{-1}$ 
\\[6pt] \hline \\[-6pt]
 SPC & $275$ & $ 993.3\pm   0.2$ & $-43.005\pm 0.003$ & $1070.0\pm   0.1$ & $-42.470\pm 0.001$\\
 ~ & $300$ & $ 977.1\pm   0.1$ & $-41.534\pm 0.002$   & $1050.2\pm   0.1$ & $-40.951\pm 0.001$\\
 ~ & $325$ & $ 956.7\pm   0.1$ & $-40.096\pm 0.001$   & $1026.5\pm   0.1$ & $-39.472\pm 0.001$\\
 ~ & $350$ & $ 933.8\pm   0.1$ & $-38.687\pm 0.002$   & $ 999.5\pm   0.1$ & $-38.019\pm 0.001$\\
 ~ & $375$ & $ 907.8\pm   0.1$ & $-37.277\pm 0.003$   & $ 969.2\pm   0.1$ &
  $-36.571\pm 0.001$\\[6pt]
 SPCE & $275$ & $1009.5\pm   0.2$ & $-48.148\pm 0.004$  & $1089.9\pm   0.1$ & $-47.594\pm 0.003$\\
    ~ & $300$ & $ 998.6\pm   0.2$ & $-46.572\pm 0.002$  & $1076.1\pm   0.1$ & $-45.968\pm 0.001$\\
    ~ & $325$ & $ 983.8\pm   0.1$ & $-45.066\pm 0.001$  & $1057.5\pm   0.1$ & $-44.408\pm 0.002$\\
    ~ & $350$ & $ 965.3\pm   0.2$ & $-43.598\pm 0.002$  & $1035.7\pm   0.1$ & $-42.897\pm 0.001$\\
    ~ & $375$ & $ 944.3\pm   0.2$ & $-42.154\pm 0.002$  & $1010.2\pm   0.1$ & $-41.405\pm 0.002$\\[6pt]
 SPCE (500 Mol.)  & $275$ & $1009.0\pm   0.1$ & $-48.147\pm 0.003$ \\
 ~ & $300$ & $ 998.4\pm   0.1$ & $-46.576\pm 0.001$ \\
 ~ & $325$ & $ 983.3\pm   0.1$ & $-45.062\pm 0.001$ \\
 ~ & $350$ & $ 964.9\pm   0.1$ & $-43.589\pm 0.001$ \\
 ~ & $375$ & $ 943.8\pm   0.1$ & $-42.142\pm 0.002$ \\[6pt]
 SPCE (864 Mol.) & $275$ & $1009.0\pm   0.1$ & $-48.152\pm 0.002$ \\
 ~ & $300$ & $ 998.2\pm   0.1$ & $-46.577\pm 0.001$ \\
 ~ & $325$ & $ 983.3\pm   0.1$ & $-45.066\pm 0.002$ \\
 ~ & $350$ & $ 964.8\pm   0.1$ & $-43.592\pm 0.001$ \\
 ~ & $375$ & $ 943.6\pm   0.1$ & $-42.149\pm 0.001$ \\[6pt]
 TIP3P & $275$ & $1005.2\pm   0.1$ & $-41.300\pm 0.002$ & $1079.8\pm   0.1$ & $-40.753\pm 0.001$ \\
 ~ & $300$ & $ 984.6\pm   0.1$ & $-39.921\pm 0.001$     & $1055.8\pm   0.0$ & $-39.329\pm 0.001$ \\
 ~ & $325$ & $ 960.9\pm   0.1$ & $-38.565\pm 0.002$     & $1028.4\pm   0.0$ & $-37.937\pm 0.001$ \\
 ~ & $350$ & $ 934.6\pm   0.2$ & $-37.229\pm 0.002$     & $ 997.7\pm   0.1$ & $-36.560\pm 0.001$ \\
 ~ & $375$ & $ 905.4\pm   0.2$ & $-35.888\pm 0.002$     & $ 963.9\pm   0.1$ &  $-35.182\pm 0.001$ \\[6pt]
TIP4P & $275$ & $1005.3\pm   0.1$ & $-42.839\pm 0.003$ & $1089.0\pm   0.1$ & $-42.957\pm 0.002$ \\
 ~ & $300$ & $ 993.5\pm   0.1$ & $-41.237\pm 0.003$    & $1075.2\pm   0.1$ & $-41.249\pm 0.001$ \\
 ~ & $325$ & $ 976.5\pm   0.1$ & $-39.697\pm 0.002$    & $1055.0\pm   0.1$ & $-39.624\pm 0.001$ \\
 ~ & $350$ & $ 955.1\pm   0.1$ & $-38.206\pm 0.002$    & $1029.9\pm   0.0$ & $-38.051\pm 0.001$ \\
 ~ & $375$ & $ 929.4\pm   0.1$ & $-36.726\pm 0.003$    & $1000.5\pm   0.1$ & $-36.508\pm 0.001$ \\[6pt]
TIP5P & $275$ & $ 987.8\pm   0.2$ & $-42.744\pm 0.009$ & $1069.5\pm   0.1$ & $-42.379\pm 0.004$\\
 ~ & $300$ & $ 982.6\pm   0.2$ & $-40.132\pm 0.004$    & $1058.6\pm   0.1$ & $-39.650\pm 0.002$\\
 ~ & $325$ & $ 964.1\pm   0.1$ & $-37.910\pm 0.003$    & $1034.2\pm   0.1$ & $-37.314\pm 0.002$\\
 ~ & $350$ & $ 936.6\pm   0.1$ & $-35.887\pm 0.005$    & $ 998.5\pm  0.1$ & $-35.248\pm 0.002$\\
 ~ & $375$ & $ 902.9\pm   0.1$ & $-33.995\pm 0.002$    & $ 960.3\pm   0.1$ & $-33.258\pm 0.002$
\\[6pt] \hline\hline
  \end{tabular}
  \caption{\footnotesize
    Average densities $\rho$ and configurational energies $E$ (per
    molecule)
    describing the examined statepoints at $p\!=\!0.1\,\mbox{MPa}$. 
    Left columns: pure water simulations. The system size is
    256 molecules except for the SPCE simulations indicated.
    Each statepoint has been simulated for $20\,\mbox{ns}$.
    Right columns: The system size is 500 water molecules plus 8 Xenon atoms.
    All systems were simulated for $60\,\mbox{ns}$.}
  \label{tab:sim}
\end{table*}
 \fi

Concerning the model parameters 
we would like to emphasise that the parameters
describing the noble gases (see Table \ref{tab:models} for details)
used here
were fitted to reproduce their pure component properties. 
Hence a perfect matching of the solubilities with the experimental data
cannot be expected.
However, since we are interested in comparing different water models,
taking these parameters is the preferred procedure since they should
work for all models equally good (or bad).
The water/gas cross terms were obtained applying the standard 
Lorentz-Berthelot mixing rules according to
$\sigma_{ij}\!=\!\left(\sigma_{ii}+\sigma_{jj}\right)/2$ and
$\epsilon_{ij}\!=\!\sqrt{\epsilon_{ii}\epsilon_{jj}}$.

\subsection{Infinite dilution properties}

Usually the solubility of a solute is measured by the Ostwald
coefficient $L^{l/g}\!=\!\rho^l_B/\rho^g_B$, where $\rho^l_B$ and
$\rho^g_B$ are the number densities of the solute in the
liquid and the gas phase, respectively, when both
phases are in equilibrium. Here $A$ denotes the solvent and $B$ indicates
the solute.
Equilibrium between both phases leads to a new expression
for $L^{l/g}$, namely,
\begin{equation}
L^{l/g}=\exp\left[ -\beta\,(\mu_{ex,B}^l-\mu_{ex,B}^g)\right],
\end{equation}
where $\beta\!=\!1/kT$ and $\mu_{ex,B}^l$ and $\mu_{ex,B}^g$ denote the excess chemical
potentials of the solute in the liquid and the gas phase, respectively.
When the gas phase has a sufficiently low density, then
$\mu_{ex,B}^g\approx 0$, hence 
$L^{l/g}$ becomes identical to the solubility
parameter $\gamma^l_B=\exp\left[ -\beta\,\mu_{ex,B}^l\right]$. For our
study, covering the temperature range between $275\,\mbox{K}$ and
$375\,\mbox{K}$, the excess chemical potential of apolar solutes
in the gas phase can be practically considered to be zero
(see Table 3 in Ref \cite{Guillot:93}).
The chemical potential of a solute  can be obtained from
a constant pressure simulation (NPT-Ensemble) 
of the pure solvent using the Widom particle
insertion method \cite{Widom:63,FrenkelSmit} according to
\begin{eqnarray}
  \label{eq:muexdef}
  \mu_B^l   
  & = &
  -\beta^{-1}\ln\frac{\left<V\right>}{\Lambda^3} \nonumber\\
  &  &
  - \beta^{-1} 
  \ln \frac{\left< V \int d\vec{s}_{N+1} \exp(-\beta\,\Delta
  U\right>}{\left<V\right>} \nonumber\\[6pt]
     & = &
  \mu_{id,B}^l\left(\left<\rho^l_B\right>\right) + 
  \mu_{ex,B}^l
\end{eqnarray}
where 
$\Delta U\!=\! U(\vec{s}^{N+1};L)-U(\vec{s}^{N};L)$ 
is the
potential energy of a randomly inserted solute $(N+1)$- particle into
a configuration containing $N$ solvent
molecules. The $\vec{s}_{i}=L^{-1}\vec{r}_{i}$ (with $L\!=\!V^{1/3}$ being the
length of a hypothetical cubic box) are the 
scaled coordinates of the particle positions and $\int\vec{s}_{N+1}$ denotes
an integration over the whole space. 
The brackets $\left<\ldots\right>$ indicate isothermal-isobaric
averaging over the configuration space of the $N$-particle system (the solvent).
$\Lambda$ represents
the thermal wavelength of the solute particle.
The first term $\mu_{id,B}^l$ is the ideal gas contribution 
of the solute chemical potential at an average solute number density 
$\left<\rho^l_B\right>=1/\left< V\right>$ at the statepoint $(T,P)$.
We would like to point out that the definition of the $\mu_{id}$ in
Eq. \ref{eq:muexdef} assumes that solute and solvent particles are of
different type and hence distinguishable.
Since we are considering water at relatively low temperatures,
the volume fluctuations are comparably small. Hence the obtained values for
$\mu_{ex}\equiv\mu^l_{ex,B}$ are
practically identical to the values obtained from constant volume simulations
at the same statepoints with differences due to the fluctuating volume 
$<0.02\,\mbox{kJ}\,\mbox{mol}^{-1}$ fore the temperature range considered 
in our study.
The entropic and enthalpic contributions to the excess chemical potential can
be obtained straightforwardly as temperature derivative according to
\begin{equation}
s_{ex}  =  - \left(\frac{\partial \mu_{ex}}{\partial T}\right)_P
\hspace*{2em}\mbox{and}\hspace*{2em}
h_{ex} =   \mu_{ex}+T\,s_{ex}
\end{equation}
and the isobaric heat capacity contribution according to
\begin{eqnarray}
c_{P,ex} & = &  -T \left( \frac{\partial^2\mu_{ex}}{\partial T^2}\right)_P\;.
\end{eqnarray}

As an alternative to the Ostwald coefficient, the solubility of gases is
often expressed in terms of Henry's constant $k_H$. 
The relationship between
Henry's constant and the solubility parameter $\gamma^l_B$ in the liquid phase is
given by \cite{Kennan:90}
\begin{equation}
k_H = \rho^l_A R T / \gamma^l_B\;,
\end{equation}
where $\rho^l_A$ is the number density of the solvent. We use this relation
in order to compare the experimental with the simulation data.

The thermodynamic solvation properties discussed in this paper
belong to the so called {\em number density
scale}. In the experimental literature 
\cite{Wilhelm:77,Rettich:81,Naghibi:86,Braibanti:94}, however, the properties
are often discussed on the {\em mole fraction scale} with the solvation free energy being
\begin{equation}
\Delta G = - R T \ln k_H\;,
\end{equation}
where Henry's constant $k_H$ is expressed in bars \cite{Ben-Naim:78,Kennan:90}. 
When comparing with
experimental data, care must been taken to which scale the discussed
properties (solvation enthalpies, entropies and heat capacities) belong.
Thermodynamic properties determined
on the mole fraction scale contain additional
terms depending on the thermal expansivity of the liquid \cite{Ben-Naim:78}.

In order to perform the calculation most efficiently we have made use of
the excluded volume map (EVM) technique \cite{Deitrick:89,Deitrick:92} by
mapping the occupied volume onto a grid of approximately $0.2\,\mbox{\AA}$ mesh-width. Distances smaller
than $0.7\times\sigma_{ij}$ with respect to any solute molecule (oxygen site)
were neglected and and the term $exp(-\beta\,\Delta U)$ taken to
be zero. With this setup
the systematic error was estimated to be less
than $0.02\,\mbox{kJ}\,\mbox{mol}^{-1}$. Although the construction of the
excluded volume list needs a little
additional computational effort, 
this simple scheme improves the efficiency of the sampling       
by almost two orders of magnitude.
For the calculation of the Lennard-Jones insertion energies
$\Delta U$ we have used cut-off distances of $9\,\mbox{\AA}$ 
in combination with a proper cut-off correction. Each configuration
has been probed by $10^3$ {\em successful} insertions (i.e. insertions into the
free volume contributing non-vanishing Boltzmann-factors).

For the case of Xenon we would also like  discuss the effect of having
a polarisable solute.
Therefore we calculate as well an additional polarisation term according to
\begin{eqnarray}
\Delta U & = & \Delta U_{LJ} + \Delta U_{pol} 
\end{eqnarray} 
with
\begin{eqnarray}
\Delta U_{pol} = -\frac{1}{2}\alpha |\vec{F}|^2\;, 
\end{eqnarray} 
where $\alpha\!=\!4.11\,\mbox{\AA}^3$ is the Xenon polarisability and
$\vec{F}$ is the electric field created by all water molecules at the
location where the particle is inserted. $\vec{F}$ is evaluated using the
classical Ewald summation technique with a Ewald convergence factor
of $2.98\,\mbox{nm}^{-1}$ 
(corresponding to a relative accuracy of the Ewald sum of $\approx10^{-4}$)
in combination with a real space cut-off of
$9\,\mbox{\AA}$ and a reciprocal lattice cut-off of $|\vec{k}_{max}|^2\!=\!25$.

We have tested our calculations by recalculating the chemical potential for
various noble gases and Methane at exactly the same 
statepoints as reported by Guillot and
Guissani \cite{Guillot:93} while explicitly taking the solute polarisability into
account. For higher 
temperatures ($>473\,\mbox{K}$) we can quantitatively 
reproduce their data, whereas for the lower temperatures 
(the temperature range of our study)
and larger particles
(Methane, Xenon)
differences occur, but which are qualitatively in accordance taking the
the estimated error of their relatively short calculations into account.

When studying particularly dense liquids, 
the Widom methods is known to fail \cite{FrenkelSmit}. 
In order to confirm the applicability of the Widom method 
we have performed additional checks on the accuracy of the
obtained chemical potentials
by comparing with results according to the
overlapping distribution method
(see section \ref{sec:Appendix} for details).

\subsection{Hydrophobic Interaction}

We use simulations containing 500 Water molecules and 8 Xenon particles to
study the temperature dependence of the association behaviour of Xenon.
The hydrophobic interaction between the dissolved Xenon particles
is quantified in terms the hydrophobic cavity potential $w(r)$ \cite{Ben-Naim:79}.
The $w(r)$ is obtained by inverting the Xenon-Xenon radial distribution functions $g(r)$, to
get the potential of mean force (PMF),
and subtracting the Xenon-Xenon pair interaction potential
\begin{equation}
w(r)=-kT \ln g(r) - V_{\rm Xe-Xe}(r)\;.
\end{equation}
We use temperature derivatives of quadratic fits of $w(r,T)$ 
to calculate
the enthalpic and entropic contributions to at each Xenon-Xenon separation
$r$. For the fits all five temperatures 
$275\,\mbox{K}$, $300\,\mbox{K}$, $325\,\mbox{K}$, $350\,\mbox{K}$
and $375\,\mbox{K}$ were taken into account.
The entropy and enthalpy contributions are then obtained as 
\begin{equation}
s(r)=-\left(\frac{\partial w(r,T)}{\partial T}\right)_P
\end{equation}
and
\begin{equation}
h(r)=w(r)+T s(r) \;.
\end{equation}
In addition, the corresponding heat capacity change relative to the bulk 
liquid is vailable according to
\begin{equation}
c_P(r) = -T \left( \frac{\partial^2 w(r,T)}{\partial^2 T} \right)_P\;.
\end{equation}

%%% Local Variables:
%%% mode: latex
%%% TeX-master: "paper"
%%% End:

\section{RESULTS AND DISCUSSION}

\subsection{Density-Curves}

\ifpretty
\begin{table*}[!ht]
  \centering
  \renewcommand{\arraystretch}{1.1}
  \ifpretty
  \renewcommand{\tabcolsep}{0.82cm}
  \else
  \renewcommand{\tabcolsep}{0.2cm}
  \fi
  \small
  \begin{tabular}{lcccc} \\ \hline\hline \\[-6pt]
  Model &
  $\rho_{0}/\mbox{kg}\,\mbox{m}^{-3}$ &
  $\rho_{1}/\mbox{kg}\,\mbox{m}^{-3}\,\mbox{K}^{-1}$ &
  $\rho_{2}/\mbox{kg}\,\mbox{m}^{-3}\,\mbox{K}^{-2}$ &
  $\rho_{3}/\mbox{kg}\,\mbox{m}^{-3}\,\mbox{K}^{-3}$ 
\\[6pt] \hline \\[-6pt]

SPCE  &  $0.65860\;10^3$  & $0.34537\;10^1$  & $-0.99533\;10^{-2}$  & $0.73989\;10^{-5}$  \\
SPC   &  $0.78387\;10^3$  & $0.25154\;10^1$  & $-0.78904\;10^{-2}$  & $0.55036\;10^{-5}$  \\
TIP3P &  $0.99460\;10^3$  & $0.95254\;10^0$  & $-0.37325\;10^{-2}$  & $0.14878\;10^{-5}$  \\
TIP4P &  $0.66286\;10^3$  & $0.32444\;10^1$  & $-0.86816\;10^{-2}$  & $0.51335\;10^{-5}$  \\
TIP5P & $-0.83369\;10^3$  & $0.16033\;10^2$  & $-0.44692\;10^{-1}$  & $0.38097\;10^{-4}$  
\\[6pt] \hline\hline
  \end{tabular}
  \caption{\footnotesize Polynomial fits of the densities obtained 
    for the pure water simulation series at $0.1\,\mbox{MPa}$ pressure:
   $\rho(T)\!=\!\rho_{0}+\rho_{1}\,T+\rho_{2}\,T^2+\rho_{3}\,T^3$.
 }
  \label{tab:dens}
\end{table*}
 
\fi
The simulated statepoints are summarised in Table \ref{tab:sim}.
Here the obtained average densities and potential energies are given.
In addition, cubic fits of the density with respect to temperature were
performed and the obtained fitting-parameters, 
which will be used later for the evaluation
of the hydrophobic solvation properties,
are listed in Table \ref{tab:dens}. The density-fits, as well as the
original data are shown in Figure \ref{fig:n01}.
A rather striking difference between the 
water models is course the
location of the density maximum. The TIP5P model
was actually parameterised to yield a density maximum at exactly
the experimental temperature and density \cite{Mahoney:2000}. 
However, Mahoney and Jorgensen used a cutoff of 1.2 nm for the water-water
interaction without any corrections. Since we apply the Ewald technique
summation here, we obtain densities which 
are consistently about $2\%$ smaller than
the values reported by Mahoney and Jorgensen, which is, however,
in accord with the observations of
Lisal et al. \cite{Lisal:2002}.

\ifpretty
\begin{figure}[!b]
  \centering
  \includegraphics[angle=0,width=7.5cm]{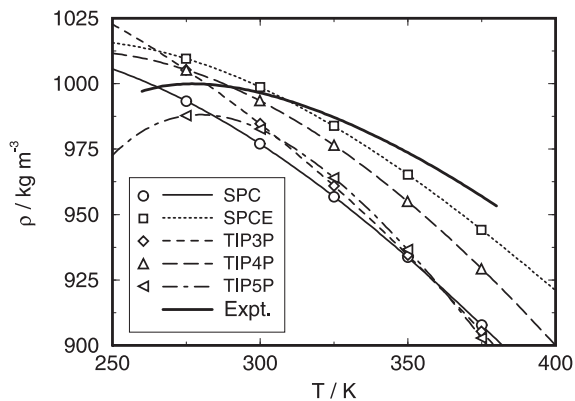}
  \caption{\footnotesize Density of water as a function of temperature for the
  different water models employed in the present study.
  The thick solid line represents experimental
  data according to Wagner and Pru{\ss}
  \cite{Wagner:2002}. 
  The thin lines represent cubic fits to the
  simulated data (see table \ref{tab:dens}
  for the fitting-parameters).}
  \label{fig:n01}
\end{figure}
 
\fi
We would like to point out
that our density-fits, although obtained at much higher temperatures,
reproduce almost quantitatively the location of the density maxima of SPCE water
\cite{Baez:94,Harrington:97,Arbuckle:2002} and TIP4P water \cite{Tanaka:96,Mahoney:2000}.
The density maxima for SPC and TIP3P 
are perhaps shifted to even lower
temperatures \cite{Billeter:94,Mahoney:2000} and thus might even lie below a possible
glass transition \cite{Sciortino:97}. 

We would like to emphasise that in line  with the temperature
dependence of the density, the different models 
show the same hierarchy 
with respect to their ability
to reproduce waters
structural features 
as shown by the work of Head-Gordon et al.
\cite{HeadGordon:2002,Hura:2000,Sorenson:2000}.
At ambient conditions the
TIP5P model agrees  quantitatively with experimental
the O-O-structure factor, SPCE and TIP4P
agree well, whereas SPC and TIP3P are completely lacking the second
peak in the O-O pair correlation function \cite{Hura:2000,Sorenson:2000}. Hence both structure and
density suggest consistently that
TIP4P and SPCE might
be considered to reflect states of 
water perhaps at {\em slightly elevated}
temperature, whereas TIP3P and SPC correspond (structurally) to states
of water at {\em much higher} temperature.
It should be mentioned that all models discussed here show a substantial
disagreement with experiment with respect to the OH and HH correlations \cite{HeadGordon:2002}.
This might be partially attributed to the fact that the models used here
are rigid. But, since as well flexible and polarisable models as well as
recent Car-Parinello simulations show the same tendency \cite{HeadGordon:2002},
the reason for this is at present not clear.

Both, solvent structure and density are considered to be of importance for the 
hydrophobic effects. Since a clear hierarchy of the quality of the water
models compared with real water is observed with respect to those properties,
the hydrophobic hydration and interaction properties might as well be affected in
such a systematic way.

%%% Local Variables:
%%% mode: latex
%%% TeX-master: "paper"
%%% End:

\subsection{Hydrophobic Hydration} 

\ifpretty\begin{figure}[!b]
  \centering
  \includegraphics[angle=0,width=7.0cm]{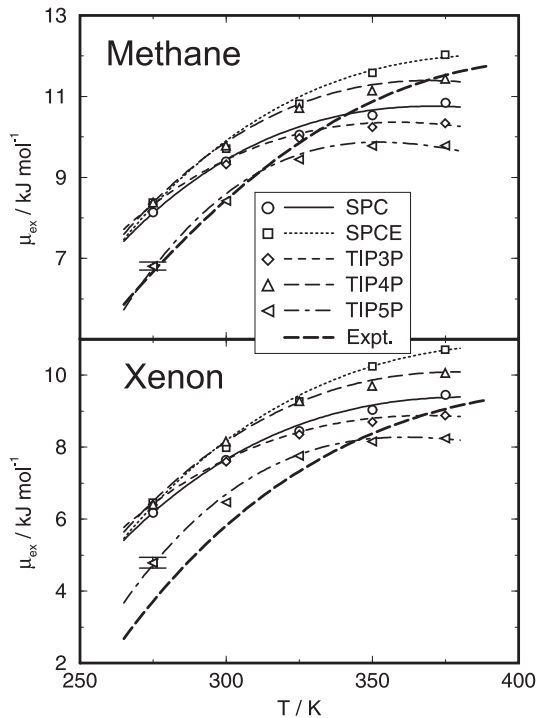}
  \caption{\footnotesize Excess chemical potential of Methane and Xenon in water as a
  function of temperature. The experimental data (heavy dashed line) are
  according to Refs. \cite{Prini:89,Wagner:2002}. The thin 
  lines represent fits to
  the simulated data employing the MIT-model.}
  \label{fig:n02}
\end{figure}
 \fi
The calculated excess chemical potentials for the noble gases
 and Methane are shown in 
Figure \ref{fig:n02} and are given in Table \ref{tab:muex}. 
The experimental data for Xenon and Methane shown in Figure \ref{fig:n02} have
been calculated using Henry's constants according to
Fernandez-Prini and Crovetto \cite{Prini:89} employing
the water
densities for the $0.1\,\mbox{MPa}$ isobar reported by Wagner and Pru{\ss} \cite{Wagner:2002}.
The thin lines in Figure \ref{fig:n02} represent fits of the data to the modified 
information theory (MIT) model, discussed in section \ref{sec:MIT}. The
 corresponding fit parameters are given in Table \ref{tab:MIT}.

The errors for the excess chemical 
potentials were calculated
using the method of Flyvbjerg and Peterson \cite{Flyvbjerg:89}
and are indicated in Table \ref{tab:muex}. A further check on the 
consistency and accuracy of the data has been  performed by application
of the method of overlapping distribution 
functions discussed in section \ref{sec:Appendix}.
We would like
to point out that the $\mu_{ex}$ data for polarisable Xenon obtained here
do only qualitatively agree with the data of
Guillot and Guissani \cite{Guillot:93}. However, test-calculations 
(results not shown here) performed for
selected statepoints given by
Guillot and Guissani 
at higher temperatures \cite{Guillot:93} do
quantitatively agree, suggesting that the differences at lower temperatures
are due to the relatively
large statistical error in their calculations and have to be attributed
to their rather short simulation runs.
Moreover, the obtained value for the excess chemical potential of Methane 
in TIP4P-water at
$300\,\mbox{K}$ of
$(9.78\pm0.1)\,\mbox{kJ}\,\mbox{mol}^{-1}$ agrees well with 
the value of $(9.79\pm0.21)\,\mbox{kJ}\,\mbox{mol}^{-1}$ obtained by Shimizu and
Chan \cite{Shimizu:2000} for $298\,\mbox{K}$ and $1\,\mbox{atm}$.

\ifpretty\begin{table*}[!ht]
  \centering
  \ifpretty
  \renewcommand{\tabcolsep}{0.42cm}
  \renewcommand{\arraystretch}{1.0}
  \else
  \renewcommand{\tabcolsep}{0.12cm}
  \renewcommand{\arraystretch}{0.82}
  \fi
  \footnotesize
  \begin{tabular}{lcccccccc|c}  
\\ \hline\hline \\[-6pt]
 ~ & ~ & \multicolumn{7}{c}{$\mu_{ex}/\mbox{kJ}\,\mbox{mol}^{-1}$} &
 $\Delta \mu_{ex}/\mbox{kJ}\,\mbox{mol}^{-1}$
\\[6pt] 
  Model &
  $T/\mbox{K}$ & 
  Ne & Ar & Kr & Xe & $\mbox{Xe}^\dagger$ & $\mbox{Xe}^*$ & $\mbox{CH}_4$ & Xe
\\[6pt] \hline \\[-6pt]
 SPC &    275  &  10.80  &   7.64  &   6.88  &   6.18   &&    3.39 &  8.14  &  -2.59  \\
   ~ &    300  &  11.41  &   8.68  &   8.12  &   7.65   &&    4.68 &  9.39  &  -3.03 \\
   ~ &    325  &  11.76  &   9.32  &   8.86  &   8.44   &&    5.56 & 10.06  &  -3.61 \\
   ~ &    350  &  11.96  &   9.77  &   9.38  &   9.04   &&    6.22 & 10.54  &  -3.95 \\
   ~ &    375  &  12.01  &  10.07  &   9.75  &   9.45   &&    6.71 & 10.84  &  -4.35 \\[6pt]
 SPCE &    275  &  10.95  &   7.76  &   7.06  &   6.46  &&   3.59 &  8.38  & -2.21   \\
    ~ &    300  &  11.73  &   8.97  &   8.41  &   7.99  &&   5.11 &  9.71  & -2.82   \\
    ~ &    325  &  12.31  &   9.94  &   9.54  &   9.27  &&   6.08 & 10.82  & -3.24   \\
    ~ &    350  &  12.70  &  10.61  &  10.37  &  10.24  &&   7.32 & 11.58  & -3.71   \\
    ~ &    375  &  12.90  &  11.07  &  10.86  &  10.71  &&   7.74 & 12.03  & -4.25   \\[6pt]
 SPCE (500 Mol.) &    275  &  10.89  &   7.66  &   6.92  &   6.15 & 6.16 &   3.36 &  8.18  \\
               ~ &    300  &  11.69  &   8.87  &   8.29  &   7.71 & 7.72 &   5.00 &  9.56  \\
               ~ &    325  &  12.27  &   9.84  &   9.41  &   9.05 & 8.96 &   6.22 & 10.67  \\
               ~ &    350  &  12.64  &  10.52  &  10.17  &   9.87 & 9.88 &   7.03 & 11.39  \\
               ~ &    375  &  12.84  &  10.94  &  10.66  &  10.41 & 10.42&   7.56 & 11.83  \\[6pt]
 SPCE (864 Mol.) &    275  &  10.88  &   7.62  &   6.80  &   6.01 & &   3.66 &  8.09  \\
               ~ &    300  &  11.65  &   8.83  &   8.20  &   7.58 & &   4.87 &  9.45  \\
               ~ &    325  &  12.24  &   9.75  &   9.24  &   8.74 & &   6.11 & 10.54  \\
               ~ &    350  &  12.61  &  10.44  &  10.05  &   9.65 & &   6.91 & 11.26  \\
               ~ &    375  &  12.82  &  10.91  &  10.63  &  10.34 & &   7.58 & 11.78  \\[6pt]
 TIP3P &    275  &  10.94  &   7.85  &   7.10  &   6.40  &&  2.71 &   8.34  & -2.81  \\
     ~ &    300  &  11.39  &   8.69  &   8.14  &   7.60  &&  4.09 &   9.33  & -3.30  \\
     ~ &    325  &  11.63  &   9.25  &   8.79  &   8.35  &&  4.85 &   9.97  & -3.74  \\
     ~ &    350  &  11.72  &   9.54  &   9.11  &   8.69  &&  5.37 &  10.24  & -4.00  \\
     ~ &    375  &  11.65  &   9.66  &   9.28  &   8.88  &&  5.77 &  10.34  & -4.32  \\[6pt]
 TIP4P &    275  &  10.88  &   7.77  &   7.09  &   6.40  &&  4.06 &   8.38  & -2.52  \\
     ~ &    300  &  11.64  &   8.99  &   8.49  &   8.16  &&  5.82 &   9.78  & -3.13  \\
     ~ &    325  &  12.14  &   9.81  &   9.44  &   9.28  &&  6.67 &  10.71  & -3.54  \\
     ~ &    350  &  12.39  &  10.31  &   9.97  &   9.70  &&  7.17 &  11.14  & -4.16  \\
     ~ &    375  &  12.44  &  10.59  &  10.31  &  10.06  &&  7.53 &  11.44  & -4.43  \\[6pt]
 TIP5P &    275  &   9.76  &   6.36  &   5.53  &   4.79  &&  2.00 &  6.81  & -1.36  \\
     ~ &    300  &  10.79  &   7.84  &   7.19  &   6.47  &&  3.34 &  8.42  & -2.46 \\
     ~ &    325  &  11.32  &   8.76  &   8.23  &   7.76  &&  4.47 &  9.44  & -3.25 \\
     ~ &    350  &  11.45  &   9.11  &   8.62  &   8.16  &&  5.15 &  9.78  & -3.83 \\
     ~ &    375  &  11.30  &   9.16  &   8.70  &   8.25  &&  5.33 &  9.78  & -4.02 
\\[6pt] \hline\hline
  \end{tabular}
  \caption{\footnotesize Calculated excess chemical potentials $\mu_{ex}$ for
  the different solutes along the $0.1\,\mbox{MPa}$ isobar obtained by
  the particle insertion technique. The $\mbox{Xe}^*$ column
  is obtained by taking the polarisability 
  of  $\alpha\!=\!4.11\;\mbox{\AA}^3$ into account. 
  The $\mbox{Xe}^\dagger$ column contains data accordig to the overlapping
  distribution method.
  The data in the most right column represent the change of
  the chemical potential when bringing
  a particle from infinity to the distance  of the  maximum of the Xe-Xe radial distribution
  function: $\Delta \mu_{ex}=\mu_{ex}(0.42\,\mbox{nm})-\mu_{ex}(\infty)$.
  The accuracy of the data was estimated to:
  $\mbox{Ne}:\pm 0.05\,\mbox{kJ}\,\mbox{mol}^{-1}$;
  $\mbox{Ar}:\pm 0.08\,\mbox{kJ}\,\mbox{mol}^{-1}$;
  $\mbox{Kr}:\pm 0.1\,\mbox{kJ}\,\mbox{mol}^{-1}$;
  $\mbox{Xe}:\pm 0.15\,\mbox{kJ}\,\mbox{mol}^{-1}$;
  $\mbox{Xe}^*:\pm 0.25\,\mbox{kJ}\,\mbox{mol}^{-1}$;
  $\mbox{Xe}^\dagger:\pm 0.05\,\mbox{kJ}\,\mbox{mol}^{-1}$;
  $\mbox{CH}_4:\pm 0.1\,\mbox{kJ}\,\mbox{mol}^{-1}$;
  $\Delta\mu_{ex}(\mbox{Xe}):\pm 0.15\,\mbox{kJ}\,\mbox{mol}^{-1}$.
  }
  \label{tab:muex}
\end{table*}
 \fi
The SPCE simulations containing larger numbers of
molecules (500 and 864) indicate that 
the excess chemical potentials obtained from 256 water molecules runs
are subject to a systematic error of $\mu_{ex}$ due to
finite size effects. The observed system-size
dependence is presumably due to the restriction of
volume fluctuations up to a certain maximum wave-length
according to the presence of periodic boundary conditions. 
As a consequence
the values from
the 256 molecule simulations are lying systematically too high.
As shown 
in Table \ref{tab:muex} this contribution is relatively small
for Neon ($\approx0.1\,\mbox{kJ}\,\mbox{mol}^{-1}$), but increases with particle
size and leads to approximately 
$0.3\,\mbox{kJ}\,\mbox{mol}^{-1}$ and
$0.5\,\mbox{kJ}\,\mbox{mol}^{-1}$
too high excess chemical potentials for Methane and Xenon, respectively.
However, this difference is
found to be only weakly temperature dependent.
Therefore the proposition of our paper,
a focus on the temperature dependence, is not affected.
The shift with respect to the {\em true} excess chemical potential
can be expected to be in the same range for all models \cite{Siepmann:92}
since the compressibilities of the different water models have
the same order of magnitude 
(between $4.6\times 10^{-5}\,\mbox{bar}^{-1}$(SPCE)
and $6.0\times 10^{-5}\,\mbox{bar}^{-1}$(TIP3P) at $300\,\mbox{K}$).

\ifpretty\begin{figure}[!t]
  \centering
  \includegraphics[angle=0,width=7.0cm]{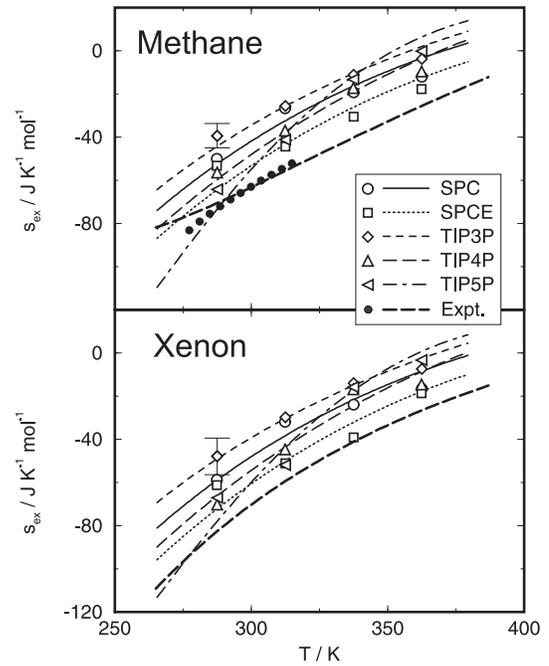}
  \caption{\footnotesize
    Entropy contribution to the excess chemical potential 
    $s_{ex}$ for
    Methane and Xenon. 
    The thick dashed lines represent the experimental data according to
    Fernandez Prini and Crovetto \cite{Prini:89}. The closed circles are
    according to the experimental data of Rettich et al. \cite{Rettich:81}.
    The employed water densities are according to Wagner and Pru{\ss} \cite{Wagner:2002}. 
    The lines are according to fits of the $\mu_{ex}$-data to the MIT model, 
    whereas the 
    symbols represent data obtained from finite differences.}
  \label{fig:n03}
\end{figure}
 \fi
Figure \ref{fig:n02} reveals that in all cases the simulated excess chemical potentials behave qualitatively like the
experimental data in the sense that the excess chemical potential is positive
and increases with temperature indicating a negative
entropy of solvation.
Please note also that around $275\,\mbox{K}$ the simulated excess chemical potentials
for all models except for the TIP5P model
 are lying rather close together, being situated
about $2.5\,\mbox{kJ}\,\mbox{mol}^{-1}$  (Methane) and 
$3.5\,\mbox{kJ}\,\mbox{mol}^{-1}$ (Xenon) above the experimental data.
With increasing temperatures, the differences between 
the values corresponding to the different water models
start to increase, already suggesting differently strong
solvation entropies. 
The overall change of $\mu_{ex}$ with temperature is smallest for
the TIP3P and SPC model, larger for TIP4P and SPCE and most extreme
for the TIP5P model. 
Please note also that for the TIP3P and TIP5P models 
the curves for Methane and Xenon suggest the presence of a
maximum of $\mu_{ex}$ close to $375\,\mbox{K}$, whereas the experimental
maximum is located at much higher temperatures.

In Figure \ref{fig:n03} the entropy contribution to the 
excess chemical potentials are shown for the different models
as well as for the experimental data. The thin lines were obtained as 
numerical derivatives of MIT-model fits whereas the symbols represent
finite differences of the $\mu_{ex}$-data given in Table \ref{tab:muex}. 
We would like to point out the the experimental data shown here
belong to the number density scale and are therefore smaller than the
values given in the paper of Rettich et al. \cite{Rettich:81},
corresponding to the mole fraction scale. 
The experimental data shown here were obtained  also 
as a numerical derivative of
the experimental excess chemical potentials
with respect to the temperature.

A detailed look at the experimental data for Methane
according to Fernandez-Prini and Crovetto \cite{Prini:89} reveals
decreased slope of $s_{ex}$ in the region below $300\,\mbox{K}$.
The reason for this is not quite clear and not present when calculating
entropies calculated from the solubility data of Rettich et al. Since this
observation
\cite{Rettich:81} is also not compatible with the calorimetric measurements
of Naghibi et al. \cite{Naghibi:86} it may probably reflect a deficiency
of the fit used in Ref. \cite{Prini:89}.
Above $300\,\mbox{K}$ both experimental 
datasets, however, do agree well.

The solvation entropy obtained for Methane in TIP4P water at 
$300\,\mbox{K}$ is found to be
$-47\pm5\,\mbox{J}\,\mbox{K}^{-1}\mbox{mol}^{-1}$,
which is slightly larger than the value of
$-40.7\,\mbox{J}\,\mbox{K}^{-1}\mbox{mol}^{-1}$
reported by Shimizu and Chan. 
The entropies are throughout negative,
in accordance with the
classical interpretation of hydrophobic hydration, suggesting
an enhanced ordering of the molecules in the solvation shell.
In the region between $275\,\mbox{K}$ and 
$300\,\mbox{K}$ the experimental data reveal about 
$15-10\,\mbox{J}\,\mbox{K}^{-1}\mbox{mol}^{-1}$
larger absolute solvation entropies for Xenon compared to Methane. 
This trend is
also observed in the simulations, where we find
$7$, 
$8$, 
$5$, 
$7$ and  
$4\,\mbox{J}\,\mbox{K}^{-1}\mbox{mol}^{-1}$ larger
absolute solvation entropies for Xenon for 
SPC, SPCE, TIP3P, TIP4P and TIP5P-water, respectively.
However, the different models show
a clear hierarchy with respect to their entropy when comparing 
with the experimental data for Methane and Xenon. For both cases, Methane and Xenon,
TIP3P and SPC exhibit the smallest solvation entropies, whereas 
TIP4P and SPCE are lying closer to the experimental data.
The TIP5P model, however, shows the strongest temperature dependence,
exhibiting the about smallest solvation entropy at $275\,\mbox{K}$ of all models
and the highest value at $375\,\mbox{K}$.
The temperature dependence of the solvation entropies can be
quantified in terms of solvation heat capacities.
The decrease of the absolute value of the solvation entropies
leads to positive solvation heat capacities (calculated for $300\,\mbox{K}$)
which are estimated to be
$(145\pm20)$, 
$(144\pm20)$, 
$(145\pm20)$, 
$(184\pm20)$, 
$(264\pm20)\,\mbox{J}\,\mbox{K}^{-1}\mbox{mol}^{-1}$ for Methane
and
$(156\pm20)$, 
$(164\pm20)$, 
$(160\pm20)$, 
$(223\pm20)$, 
$(310\pm20)\,\mbox{J}\,\mbox{K}^{-1}\mbox{mol}^{-1}$ for Xenon
in SPC-, SPCE-, TIP3P-, TIP4P- and TIP5P-water, respectively.
The experimental data of $c_{P,ex}$ according to the solubility data
of Methane of Rettich et al. \cite{Rettich:81} was obtained to
be $234\,\mbox{J}\,\mbox{K}^{-1}\mbox{mol}^{-1}$ at $300\,\mbox{K}$,
when transforming their data on the number density scale.
This value has also
been reported by Widom et al. \cite{Widom:2003}.
The value of $c_{P,ex}$ for Xenon according to Fernandez-Prinis
data using the same procedure
was estimated to be $280\,\mbox{J}\,\mbox{K}^{-1}\mbox{mol}^{-1}$.
The general trend, larger absolute solvation entropies and heat-capacities
for Xenon compared to Methane, is accomplished by all water models and is
also found in experiment.
Moreover, the experimental data indicate an increase of the heat capacities at lower
temperatures. The fitted curves shown in Figure \ref{fig:n03} tend to
suggest this as well. However,
the error in the entropies obtained from finite differences is perhaps too large
to definitely confirm this. The values for $c_{P,ex}$ given above
reflect
an average considering the whole temperature range.

\ifpretty\begin{figure}[!t]
  \centering
  \includegraphics[angle=0,width=7.0cm]{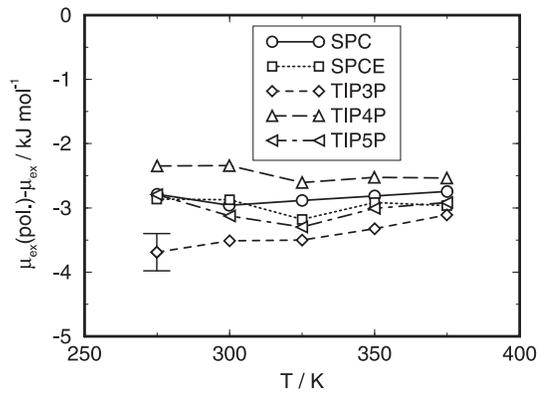}
  \caption{\footnotesize 
    Contribution of the polarisability to the excess chemical potential
    of Xenon in water.}
  \label{fig:n04}
\end{figure}
 \fi
We would like to point out that the apparent hierarchy of the models
with respect to their ability to reproduce waters 
structure (OO-correlation) and the temperature of 
maximum density is also observed when comparing the solvation entropies
of simple solutes. TIP3P has by far the smallest solvation entropy and the SPC
model is only slightly better. SPCE is apparently the best solvent model at
higher temperatures (from the point of view
of solvation entropies), whereas the TIP5P model is is closer to
the experimental data below $300\,\mbox{K}$.
Apparently the water models that exhibit a more strongly pronounced
structure are also subject to an enhanced ordering of the molecules in the
solvation shell, which is reflected by more negative hydration entropies. 
However, this is perhaps not too surprising, since a water
model that provides a better representation of waters structure might as
well lead to a more realistic structural description of the hydrophobic
hydration shell. The reason for this might be related to the observation 
that the liquid water structure contains 
cavities suitable for hydrophobic particles \cite{Geiger:86}. 
This is, of course, also the cause for the Widom particle
insertion technique performing
so well for hydrophobic particles in water.

The density-curves and corresponding structural features of the individual water models
provoked the simplistic interpretation that
TIP3P and SPC corresponds structurally to states of
water at increased temperature. We would like to point out that the simulated entropies support
this point of view. To a first order approximation the simulated entropy-data
can be simply corrected by a temperature shift.

Finally, we would like to discuss the effect of having an
explicitly polarisable solute
particle. As shown in Figure \ref{fig:n04}, introducing 
a polarisable Xenon particle
leads to a lowering of the chemical potential of about 
$2.5\,\mbox{kJ}\,\mbox{mol}^{-1}$ to
$3.5\,\mbox{kJ}\,\mbox{mol}^{-1}$. Please note that
 the contribution to the chemical
potential due to the polarisability is only weakly temperature dependent.
The most extreme cases in this respect are the TIP3P and the TIP4P model.
The TIP3P-data show
a change of about $0.5\,\mbox{kJ}\,\mbox{mol}^{-1}$ over the entire
temperature range, indicating a
lowering of the entropy for each temperature on average 
of about $5\,\mbox{J}\,\mbox{K}^{-1}\mbox{mol}^{-1}$
For the TIP4P model, however, 
this contribution is about $-2\,\mbox{J}\,\mbox{K}^{-1}\mbox{mol}^{-1}$, leading
to a slight increase of the solvation entropy.
Both values are lying in the range of
the expected error for the entropy values 
derived from finite difference values
of about $\pm8\,\mbox{J}\,\mbox{K}^{-1}\mbox{mol}^{-1}$.

%%% Local Variables:
%%% mode: latex
%%% TeX-master: "paper"
%%% End:

\subsection{A Modified Information Theory Model}

\label{sec:MIT}

\ifpretty\begin{figure}[!b]
  \centering
  \includegraphics[angle=0,width=7.0cm]{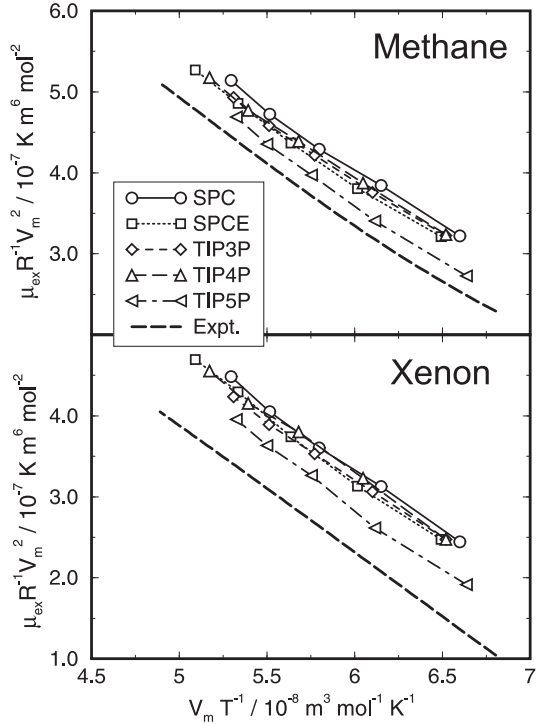}
  \caption{\footnotesize
    Scaling plots of the excess chemical potential of Methane and Xenon in
    water according to the MIT-model.
    The MIT-model parameters $c$ and $a$ were obtained as slope and
    intersection of the data shown here, assuming
    linearity between $\mu_{ex}\,V_m^2$
    and $V_m/T$.}
  \label{fig:n05}
\end{figure}
 \fi
In the previous section we discussed the properties of hydrophobic hydration
by relating structural features and solvation entropies employing (to a first
order approximation) a temperature shift. In this
section we will try to elaborate a more quantitative way to
describe the differences between the models.

\ifpretty\begin{figure}[!t]
  \centering
  \includegraphics[angle=0,width=7.0cm]{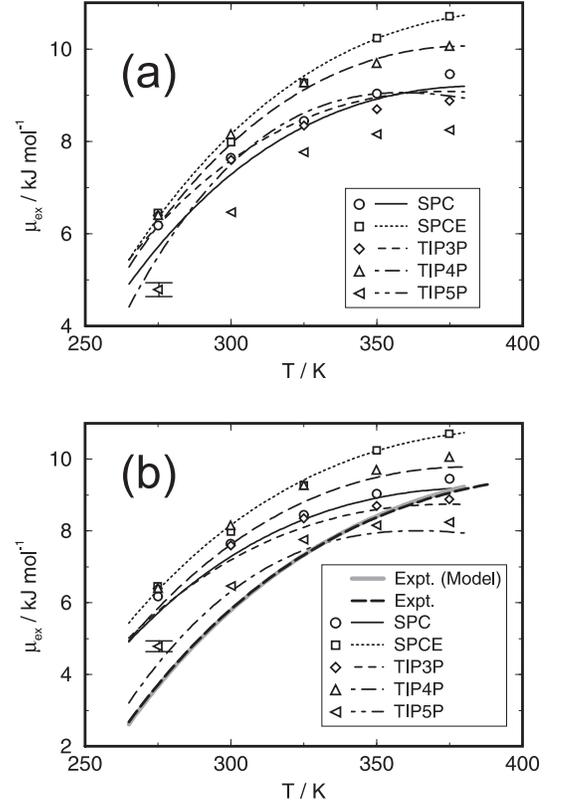}
  \caption{\footnotesize Excess chemical potential of Xenon in water as a
    function of temperature. (a): 
    The lines represent MIT-Model predictions data using
     just the parameters obtained for SPCE-water and the different
     model densities.
     (b): The lines represent MIT-Model predictions employing
     the SPCE-parameters, but scaling
     $a$ by
     $(\sigma_{\rm Xe-model}/\sigma_{\rm Xe-SPCE})^6$. 
     To predict the experimental data a hypothetical 
     ``$\sigma_{\rm Xe-model}$'' of $3.5275\;\mbox{\AA}$ has been used.}
  \label{fig:n06}
\end{figure}
 \fi
The information theory approach to the hydrophobic hydration according to
Hummer et al. \cite{Hummer:2000} suggests that the excess chemical potential of a
hard sphere particle (i.e. a spherical cavity) can be
expressed as
\begin{eqnarray}
\frac{\mu_{ex}}{kT}&\approx& \frac{v^2\rho^2}{2\sigma_n^2} \;,
\end{eqnarray}
where $v$ is the volume of the spherical cavity, $\rho$ 
is the water number density
and $\sigma_n\!=\!\left< n^2 \right>-\left< n \right>^2$ is the variance of
the occupancy number of water molecules in the sphere.
For the case of noticeable attractive interactions Hummer et al. 
\cite{Hummer:2000} propose
the presence of an additional attractive term plus an offset
according to
\begin{eqnarray}
\label{eq:ITmodel}
\frac{\mu_{ex}}{kT}&\approx& A \rho^2 - C\rho/T + B
\end{eqnarray}
where $A\!=\!v^2\rho^2/(2\sigma_n^2)$ and the
parameter $C$  is found to qualitatively account for the effects
of attractive solute-solvent interactions \cite{Hummer:2000}.
However,
in order to arrive at a quantitative description of the experimental
and simulated data, we find it advantageous to express the parameters
$A$ and $C$, both inverse proportional to the temperature with
\begin{equation}
\label{eq:acdef}
A=a'/T \;\;\;\mbox{and} \;\;\;C=c'/T \;.
\end{equation}
The justification for this approach is purely empirical
but can be perhaps rationalised as:
1.) An effect of a temperature dependent 
{\em effective} particle diameter, as well as a slight increase in 
the fluctuation
$\sigma_n^2$ \cite{Hummer:2000}. At higher temperature, the effective diameter of the 
solute is likely to decrease due to the form of the repulsive
potential. 
2.) An effect of a  more strongly weakened attractive 
interaction with increasing temperature.
Moreover, when doing this, the offset $B$ can be dropped 
($B\!=\!0$), although we must admit that the original 
model in Eq. \ref{eq:ITmodel} using three parameters
represents the data slightly better.

\ifpretty\begin{table*}[!ht]
  \centering
  \renewcommand{\arraystretch}{1.0}
  \ifpretty
  \renewcommand{\tabcolsep}{0.50cm}
  \else
  \renewcommand{\tabcolsep}{0.21cm}
  \fi
  \small
  \begin{tabular}{llcccccc|c} \\ \hline\hline \\[-6pt]
  Model &
  MIT-parameters & 
  Ne & Ar & Kr & Xe & $\mbox{Xe}^*$ & $\mbox{CH}_4$  & $\mbox{Xe}^c$
\\[6pt] \hline \\[-6pt]
  SPC & $a/10^{-6}\mbox{K}\,\mbox{m}^6\mbox{mol}^{-2}$ 
&1.130&1.169&1.213& 1.254&1.050&1.272&0.640\\
      & $c/\mbox{K}^2\mbox{m}^3\mbox{mol}^{-1}$        
&-10.7 &-13.2 &-14.3 &-15.3 &-13.9 &-14.4&-7.47 \\[6pt]
 SPCE & $a/10^{-6}\mbox{K}\,\mbox{m}^6\mbox{mol}^{-2}$ 
& 1.085& 1.159& 1.218& 1.280&1.076&1.271&0.749\\
      & $c/\mbox{K}^2\mbox{m}^3\mbox{mol}^{-1}$        
&-10.3 &-13.4 &-14.7 &-16.0 &-14.5 &-14.7&-9.03 \\[6pt]
TIP3P & $a/10^{-6}\mbox{K}\,\mbox{m}^6\mbox{mol}^{-2}$ 
& 1.121& 1.138& 1.165& 1.184& 1.002&1.230&0.569\\
      & $c/\mbox{K}^2\mbox{m}^3\mbox{mol}^{-1}$        
&-10.8 &-12.9 &-13.7 &-14.4 &-13.8 &-14.0&-6.54 \\[6pt]
TIP4P & $a/10^{-6}\mbox{K}\,\mbox{m}^6\mbox{mol}^{-2}$ 
& 1.096& 1.151& 1.190& 1.240& 1.029&1.245&0.654\\
      & $c/\mbox{K}^2\mbox{m}^3\mbox{mol}^{-1}$        
&-10.5 &-13.1 &-14.1 &-15.2 &-13.3 &-14.2&-7.61 \\[6pt]
TIP5P & $a/10^{-6}\mbox{K}\,\mbox{m}^6\mbox{mol}^{-2}$ 
& 1.133& 1.173& 1.200& 1.223& 0.972&1.256&0.475\\
      & $c/\mbox{K}^2\mbox{m}^3\mbox{mol}^{-1}$        
&-11.3 &-13.9 &-14.8 &-15.6 &-13.5 &-14.9&-5.08\\[6pt]
Expt. & $a/10^{-6}\mbox{K}\,\mbox{m}^6\mbox{mol}^{-2}$ 
& 1.104& 1.191& 1.212& 1.174& ~ &1.243\\
      & $c/\mbox{K}^2\mbox{m}^3\mbox{mol}^{-1}$        
&-10.9 &-14.2 &-15.5 &-15.7 & ~ &-15.1
\\[6pt] \hline\hline
  \end{tabular}
  \caption{\footnotesize Parameters describing the excess chemical potential of the noble
  gases and Methane applying the modified information theory (MIT) model. The
  model parameters were obtained by fitting the data of Table \ref{tab:muex}.
  The $\mbox{Xe}^*$-data corresponds to the polarizable Xe-particle.
  The $\mbox{Xe}^c$-data corresponds a Xenon particle in contact with another
  Xenon particle.
  The parameters representing the experimental data were fitted to the
  data published in Refs. \cite{Prini:89} and \cite{Wagner:2002}.}
  \label{tab:MIT}
\end{table*}
 \fi
Since this fitting procedure has been 
basically inspired by the information theory
(IT) approach for hydrophobic hydration and interaction
we will refer to it as 
{\em modified information theory model} 
(MIT) in the course of this paper. Moreover, the parameters 
$a$ and $c$ are expressed in terms of the molar volume, such that
\begin{equation}
  \label{eq:MITmodel}
  \frac{\mu_{ex}}{R}=\frac{a}{V_m^2}+\frac{c}{V_m\,T}
\end{equation}
with $R$ being the ideal gas constant.

In Figure \ref{fig:n05} we show scaled plots of the chemical potential of
Xenon and Methane according to equation \ref{eq:MITmodel}.
Taking the experimental data, the two parameter
MIT model accurately represents
the data of the noble gases up to temperatures of
$420\mbox{K}-450\mbox{K}$.
At higher temperatures the linearity between $\mu_{ex}\,V_m^2$
and $V_m/T$ breaks down, which is not unexpected
due to the enhanced increase of the isothermal compressibility $\kappa_T$
\cite{Hummer:2000}.
Comparing the obtained MIT-parameters for the noble gases
(see Table \ref{tab:MIT}), it is evident that the
parameters also behave meaningfully
when going from Ne to Xe,
in the sense that the $c$-parameters becomes more negative, 
hence 
the attractive interaction becomes stronger,
whereas the increasing $a$-parameter 
accounts for the increasing size of the hydrophobic particle.

Nevertheless, the limits of the model become evident also:
The slight deviation from linearity for Methane 
(which is fully absent in the case of Xenon)
in Figure \ref{fig:n05} becomes significantly stronger
in the case of Neon (not shown), indicating that both parameters
are subject to a enthalpy-entropy compensation 
effect 
and a sufficiently strong attractive 
(e.g. a large Lennard-Jones $\epsilon$)
interaction is required for a good performance of the model.
Or in other words: A counterbalancing of the two terms
is necessary.

Moreover, it  is observed that the $a$ parameter obtained for 
Xenon is even smaller (or at least very close) to the value obtained
for Methane. A stronger attractive interaction apparently 
leads to a shrinking apparent cavity size which is maybe related to
the presence of a more tightly bound hydration shell.
This feature is strongly pronounced in the case of the experimental values,
here the $a$-parameter for Xenon is even smaller than the
value for Krypton. This might be well attributed to the larger
polarisability of the Xenon atom compared with Krypton
and Methane. As shown in Table \ref{tab:MIT},
taking the polarisability into account significantly
reduces the $a$ and $c$.

Figure \ref{fig:n05} reveals that the 
scaled chemical potential 
data for all models (except for TIP5P) are
lying quite close to each other. In addition, the
experimental line has almost the same slope, so that the
difference between the different models (and experiment)
can be described almost completely by just
shifting the $a$-parameter. Moreover, the data according
to the different models (except for TIP5P) 
fall almost onto
one line. As a consequence it is worthwhile trying to
describe the chemical potentials by using the same set of parameters
(here we take the data corresponding to the SPCE-model), while 
just taking
the density-curves of the different models into account.

In Figure \ref{fig:n06}a the  Xenon excess chemical potentials for the
different water-models,
as well as the MIT-model predictions based on the SPCE-parameters
are shown.
An almost quantitative prediction is achieved, except for
the TIP5P model. A possible explanation 
for the large deviation of TIP5P-data
might be the noticeably  smaller
Lennard-Jones $\sigma$ of the TIP5P-model (see table \ref{tab:models}).
The information theory suggests that the $a$-parameter should scale
with the square of the the particle volume. Hence we try to improve
the model prediction by scaling the $a$-parameter with the
factor $\left(\sigma_{\rm Xe-model}/\sigma_{\rm Xe-SPCE}\right)^6$,
shown in Figure \ref{fig:n06}b, which, indeed,
leads to a substantial improvement for the case of TIP5P.
In order reproduce the experimental 
data, a scaling procedure taking the experimental
density-curve and and Lennard-Jones parameter 
of $\sigma_{\rm Xe-Water}\!=\!3.5275\;\mbox{\AA}$ has to be employed.

Apparently the suggested MIT model is able to describe the
excess chemical potential of Xennon
for the different water models and experiment
 by just taking the different density isobars
into account. In the previous section, however, it was argued that both
water structure and the
solvation entropies are well related and hence responsible for the performance
of the different models. However, this is perhaps not a contradiction when 
keeping in mind that waters structural and density changes are
tightly related and that the transformation towards a more
tetrahedrally ordered structure at lower temperatures is the basis
for the presence of a density maximum \cite{Paschek:99}. 

We are quite confident that similarities between experimental 
and simulated  data applying the suggested
rescaling procedure reveal a signature for hydrophobic hydration in the
lower temperature regime and might also be helpful when trying to quantify the
interaction between hydrophobic of particles.

%%% Local Variables:
%%% mode: latex
%%% TeX-master: "paper"
%%% End:

\subsection{Hydrophobic Interaction}

\ifpretty
\begin{figure}[!b]
  \centering
  \includegraphics[angle=0,width=7.0cm]{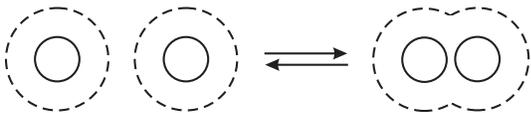}
  \caption{\footnotesize 
    Schematic diagram of the hydrophobic 
    association process according to the classic picture: The contact
    configuration is stabilised with increasing temperatures
    by minimising the entropy penalty.}
  \label{fig:n07}
\end{figure}

\begin{figure}[!t]
  \centering
  \includegraphics[angle=0,width=7.0cm]{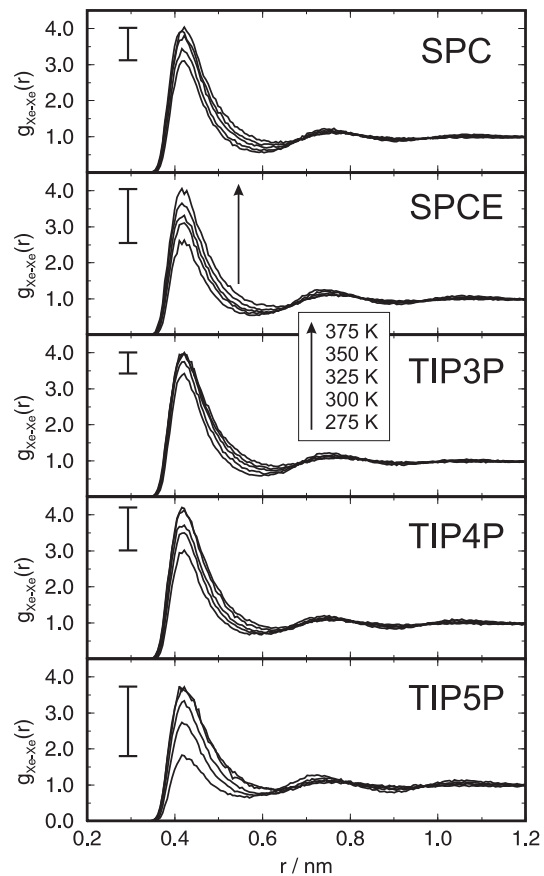}
  \caption{\footnotesize
    Xe-Xe radial pair distribution functions $g(r)$
    for the different water models and temperatures. The arrow indicates
    the sequence of $g(r)$-curves pointing 
    from low to high temperatures. The bar
    indicates the change of the height of the first 
    maximum over the whole temperature range.
  }
  \label{fig:n08}
\end{figure}

\begin{figure}[!t]
  \centering
  \includegraphics[angle=0,width=7.0cm]{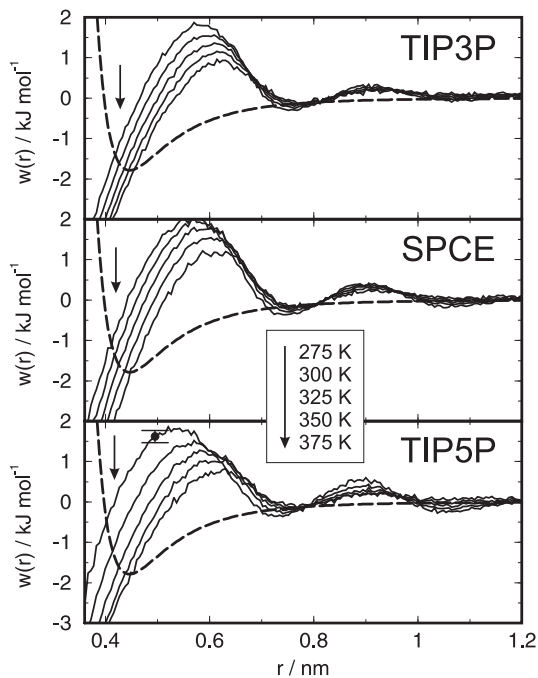}
  \caption{\footnotesize
    Cavity part of the profile of free energy $w(r)$ obtained 
    for the association of two Xenon particles
    for three selected models
    and all temperatures. 
    The heavy dashed lines denote the Xe-Xe 
    Lennard-Jones pair potential.}
  \label{fig:n09}
\end{figure}

%%% Local Variables: 
%%% mode: latex
%%% TeX-master: "paper"
%%% End: 

\fi
In order to quantify  the hydrophobic interaction we calculate the
Xenon-Xenon pair distribution functions for the different models and
temperatures. The data are shown Figure \ref{fig:n08}. All models show an increase
of the first peak with rising temperature. This observation is well in
accordance with the interpretation that
the association
of two hydrophobic particles is stabilised by minimising the solvation
entropy penalty and 
 has already been reported by a large number of publications
\cite{Smith.D:92,Smith.D:93,Dang:94,Luedemann:96,Luedemann:97,Rick:97,%
Shimizu:2000,Rick:2000,Shimizu:2001,Ghosh:2002}. 
The idea is schematically depicted in Figure \ref{fig:n07}:
The enhanced ordering of the solvent molecules in the hydration shell
results in a negative solvation entropy. If two particles associate,
the corresponding hydration shells overlap, hence leading to a positive net
entropy. 
Consequently contact-configurations should become increasingly
stabilised at higher temperatures. In parallel, the increased heat capacity
of the water molecules in the hydrophobic
hydration shell should lead to a negative heat
capacity contribution for the association of two particles, thus weakening the
the entropy contribution at elevated temperatures. 
We would like to emphasise
that this model of course lacks completely detailed
structural considerations.
Our study reveals significant
differences for the different water models.
The overall maximum variation 
of the height of the first peak (which is also indicated in
Figure \ref{fig:n08}) is smallest for the TIP3P and SPC models,
TIP4P and SPCE show a stronger variation, whereas the TIP5P model
reveals an extremely pronounced dropping of the first peak when going to
lower temperatures. In all cases
the maximum of the first peak is located at a distance of  about 
$4.2\,\mbox{\AA}$, which does not change significantly with 
varying temperature.
Moreover, the pair distributions functions reveal as well
the presence of a pronounced second peak corresponding to the solvent separated
Xenon-Xenon pair configuration, which is located at $7.2-7.8\,\mbox{\AA}$
and is shifting to smaller distances with decreasing temperatures. In
addition, a third  maximum, being located at about $10.5\,\mbox{\AA}$, seems to
exist at least at the lowest temperatures. 

\ifpretty \begin{figure}[!t]
  \centering
  \includegraphics[angle=0,width=7.0cm]{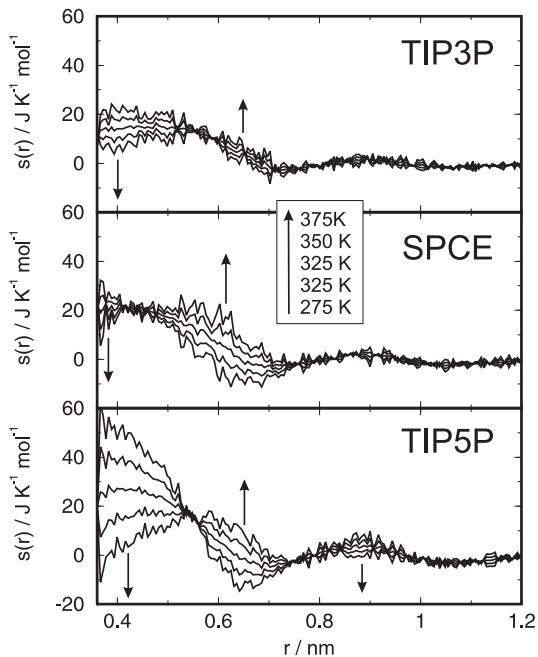}
  \caption{\footnotesize
    Entropy profile $s(r)$ for the association of two Xenon particles.}
  \label{fig:n10}
\end{figure}
 \fi
In Figure \ref{fig:n09} the corresponding profiles of free energy 
for the association of two Xenon particles are shown. 
The TIP3P-, SPCE- and TIP5P-water 
models might be taken here as the representative cases for {\em small},
{\em medium} and {\em strong} temperature variation, respectively. 
In addition, the relative change of the excess
chemical potentials when bringing a Xenon particle from the bulk to the
Xenon-Xenon contact distance 
$\Delta\mu_{ex}\!=\!\mu_{ex}(0.42\,\mbox{nm})-\mu_{ex}(\infty)$
are given in Table \ref{tab:muex} for all models and temperatures.
We would like to point
out that quadratic fits with respect to the temperature describe the
changes of data over the entire temperature range reasonably well.
From the distribution of pair correlations 
functions obtained from different parts of
the simulation runs, we conclude that the individual
curves shown here are accurate
within an interval of $\pm0.15\,\mbox{kJ}\,\mbox{mol}^{-1}$.
For completeness Figure \ref{fig:n09} contains 
as well the {\em temperature independent}
pair potential between two Xenon particles.
Figure \ref{fig:n09} shows that with decreasing temperature the contact
configuration is increasingly destabilised, whereas the solvent separated
configuration becomes more and more stable. 

\ifpretty \begin{figure}[!t]
  \centering
  \includegraphics[angle=0,width=7.0cm]{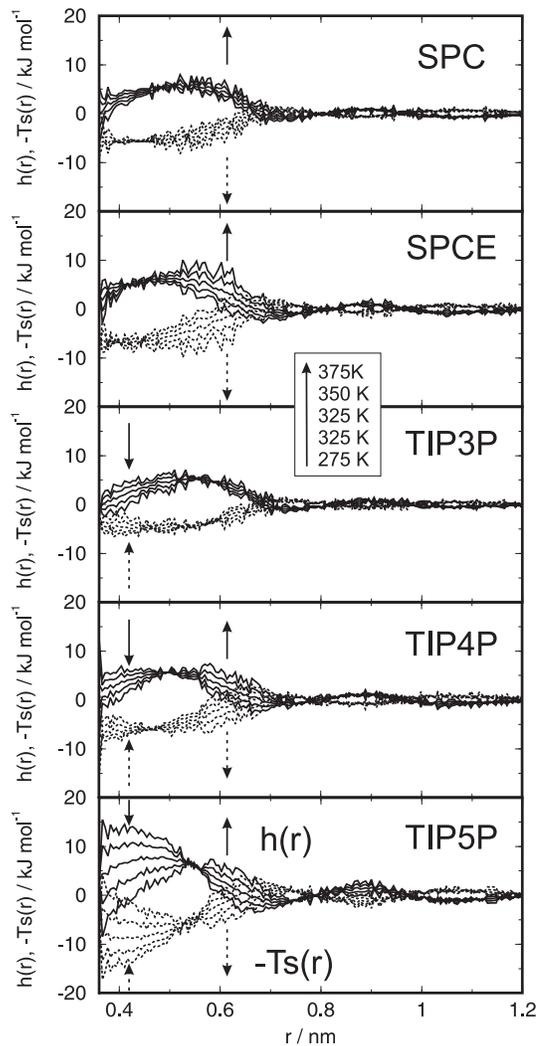}
  \caption{\footnotesize 
    Enthalpic and entropic contributions to the cavity part of the
    profile of free energy
    $w(r)=h(r)-Ts(r)$. Solid lines: $h(r)$. Dotted lines: $-Ts(r)$.}
  \label{fig:n11}
\end{figure}
 \fi
From the temperature derivative 
of the $w(r,T)$ data set we obtain the entropy-profiles for the association
of two Xenon particles, shown in
Figure \ref{fig:n10}. The entropic $-T\,s(r)$ and enthalpic $h(r)$
contributions to the profile of free energy are given in 
Figure \ref{fig:n11}. 
The temperature variation of the entropy profiles 
is quantified by the corresponding
heat capacity profiles $c_P(r)$ given in Figure \ref{fig:n14}.
As already shown by
Smith and Haymet \cite{Smith.D:92,Smith.D:93}
and others, the hydrophobic association
process is found to be entropically favoured and enthalpically disfavoured in
case of all models.
The value of $-5.3\,\mbox{kJ}\,\mbox{mol}^{-1}$ for the entropic
contribution to the profile of free energy at 
$300\,\mbox{K}$ at the contact distance for TIP3P water 
is reasonably close to the $\approx-3.5\,\mbox{kJ}\,\mbox{mol}^{-1}$ 
(when taking the data from Figure 5 in Ref. \cite{Ghosh:2002})
reported for
Methane by Ghosh et al. \cite{Ghosh:2002} and
the value of $-6.5\,\mbox{kJ}\,\mbox{mol}^{-1}$ is in
the same sense consistent with the
$-4.14\,\mbox{kJ}\,\mbox{mol}^{-1}$ for
Methane in TIP4P water obtained by Shimizu et al.\cite{Shimizu:2000}. 
The observation of slightly larger 
entropies for the association of Xenon instead of Methane is 
plausible since the hydration entropies for Xenon
are also found to be more negative.
We would also like to point out that
the hierarchy of the data for the different models according
the published data is apparently
consistent with our observations.
The data of Rick \cite{Rick:2000} 
obtained from simulations employing the polarisable TIP4P-FQ model
have been criticised
\cite{Shimizu:2000, Shimizu:2001, Ghosh:2002}
because of the unreasonably high heat capacity change for the
association of two Methane particles.
However, the large
entropy contribution to
the profile of free energy of about $-12\,\mbox{kJ}\,\mbox{mol}^{-1}$ 
at the contact distance
is rather close to the value of $-11.5\,\mbox{kJ}\,\mbox{mol}^{-1}$ 
that is observed here
for the TIP5P model.

\ifpretty \begin{figure}[!b]
  \centering
  \includegraphics[angle=0,width=6.8cm]{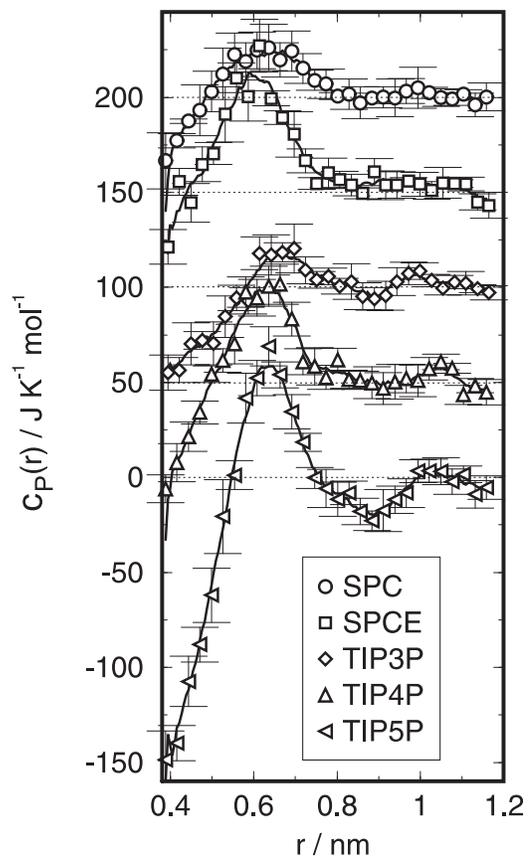}
  \caption{\footnotesize
    Relative change in heat capcacity $c_P(r)$
    for the hydrophobic interaction
    between two dissolved Xenon particles.
    The data correspond to the temperature derivative of quadratic fits
    of $w(r,T)$  obtained for $300\,\mbox{K}$.
    For displaying purposes the data are shifted by an offset of
    $50\,\mbox{J}\,\mbox{K}^{-1}\,\mbox{mol}^{-1}$.
    }
  \label{fig:n12}
\end{figure}
 \fi
Figure \ref{fig:n10} shows that
in all cases there is
a tendency for the entropy at very short distances to become smaller with increasing 
temperature, whereas in the region around $6\,\mbox{\AA}$, at the so
called {\em desolvation barrier}, the entropy
 increases with temperature.
The general tendency to smaller contact-entropies is in accordance with the
decrease of the absolute values of the solvation entropies.
However, the distances where the entropy-curves are found to cross each other are 
quite diverse. Whereas for the TIP-models this region is located
around $5.5\,\mbox{\AA}$ it is found in case of the SPC and SPCE models
at about $4.5\,\mbox{\AA}$, much closer to the Xenon-Xenon pair distance
of about $4.2\,\mbox{\AA}$. As a consequence the pair formation entropies 
vary much more strongly in case of the TIP models. 

From Figure \ref{fig:n11} it is evident that in almost all cases the
pair configuration is entropically stabilised and enthalpically
destabilised. However, TIP5P shows a much stronger 
enthalpy-entropy compensation effect than all other models,
but as well a much stronger variation with temperature. 
In addition, also
the signature of the hydrophobic association vanishes in
case of the TIP5P and TIP3P models at $375\,\mbox{K}$ where the entropic
and enthalpic contribution to the cavity potential at the the contact distance
have almost the same size. 
This is consistent with finding that in case of the TIP5P and TIP3P
models maximum of $\mu_{ex}$ for Xenon is close to $375\,\mbox{K}$ and
with the contact entropy $s_{ex}$ being close to zero.

The temperature dependence discussed here can be quantified by the association
heat capacities $c_P(r)$ shown in Figure \ref{fig:n12}. 
Shimizu and Chan \cite{Shimizu:2000,Shimizu:2001,Shimizu:2002}
report for the association of Methane particles in TIP4P
water a change of the heat capacity for the contact state close to zero in qualitative
disagreement with the approximate net effect according to the overlapping hydration shells.
In addition, they observe a maximum of the heat capacity of about 
$120\,\mbox{J}\,\mbox{K}^{-1}\,\mbox{mol}^{-1}$ at the location of the
desolvation barrier at a distance of
$5.5\,\mbox{\AA}$. In a previous study using a polarisable
water model Rick \cite{Rick:2000} did not observe any peak
at the desolvation barrier. Nevertheless, his value of about $-2500\,\,\mbox{J}\,\mbox{K}^{-1}\,\mbox{mol}^{-1}$
for the relative change of heat capacity for the contact state 
is unreasonably large, being about more than one order of
magnitude larger than the  solvation heat capacity of Methane found in
experiment and the water models used in our study. 
However, in a  more recent study Rick \cite{Rick:2003} confirms the
presence of a peak at the desolvation barrier of about
$40\,\mbox{J}\,\mbox{K}^{-1}\,\mbox{mol}^{-1}$ for the polarizable
FQ model.
In  qualitative agreement with Rick's recent study \cite{Rick:2003} and
Shimizu et al. \cite{Shimizu:2001,Shimizu:2002} 
and Southall and Dill \cite{Southall:2002:2} we find
that all models show a peak in the heat capacity at the
desolvation barrier in the region around $6\,\mbox{\AA}$ of about
$20\,\,\mbox{J}\,\mbox{K}^{-1}\,\mbox{mol}^{-1}$ for the TIP3P and SPC model,
about $70\,\,\mbox{J}\,\mbox{K}^{-1}\,\mbox{mol}^{-1}$ for the TIP5P and SPCE
model and about $50\,\,\mbox{J}\,\mbox{K}^{-1}\,\mbox{mol}^{-1}$ for the
TIP4P.

However, for the change in heat capacity for the contact state the different
models show a quite diverse behaviour: All TIP models reveal a negative
contribution to the heat capacity for the contact state 
(TIP3P: $-44\,\,\mbox{J}\,\mbox{K}^{-1}\,\mbox{mol}^{-1}$; 
TIP4P: $-42\,\,\mbox{J}\,\mbox{K}^{-1}\,\mbox{mol}^{-1}$; 
TIP5P: $-130\,\,\mbox{J}\,\mbox{K}^{-1}\,\mbox{mol}^{-1}$) with the value
for the TIP5P model being extremely large.
The Berendsen models reveal significantly smaller values with 
$-23\,\,\mbox{J}\,\mbox{K}^{-1}\,\mbox{mol}^{-1}$ for SPC and
$+5\,\,\mbox{J}\,\mbox{K}^{-1}\,\mbox{mol}^{-1}$ for the SPCE model.
Hence here we apparently observe for Xenon and the SPCE model what Shimizu and
Chan observe for Methane in TIP4P water: A slightly positive heat capacity
contribution. 
The apparent disagreement between the behaviour of Methane and
Xenon in TIP4P water, however, might be related to the detailed
hydrogen bonding situation (and their flucuations) around the differently
sized particles. We would also like to point out that
Rick \cite{Rick:2003} finds a negative heat capacity
contribution for the Methane-Methane contact pair in TIP4P water
of about $-140\pm 120\,\mbox{J}\,\mbox{K}^{-1}\,\mbox{mol}^{-1}$.

The apparently smaller heat capacity contributions of the Berendsen-models 
and the extremely large value found for the TIP5P model, however, strongly
suggest that details in the hydrogen bonding situation (i.e. structure of the
hydration shell) of the Xenon-Xenon contact state are responsible for the
differences. A further detailed comparative analysis concerning the energetics,
hydrogen bonding fluctuations and structure of the hydration shell 
for the different models will help to reveal this differences more clearly
and is the topic of a following publication.

%%% Local Variables:
%%% mode: latex
%%% TeX-master: "paper"
%%% End:

\subsection{Concerning the Hydrophobic Interaction Between Xenon Particles in Real Water}

\ifpretty
\begin{figure}[!b]
  \centering
  \includegraphics[angle=0,width=7.0cm]{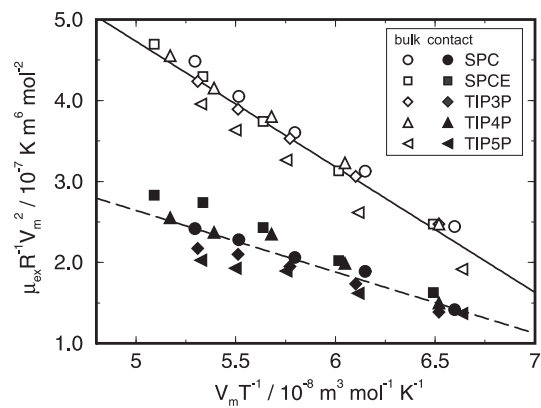}
  \caption{\footnotesize
    Scaled plots of the excess 
    chemical potential $\mu_{ex}$ of Xenon in
    water for the different water models 
    according to the MIT-model. Open symbols: chemical potential 
    obtained for the
    water bulk. Closed symbols: chemical potential in the vicinity 
    of a Xenon particle
    (obtained at the distance of the first peak of the Xe-Xe g(r)-function with
    $r\!=0.42\;\mbox{nm}$). The lines represent MIT fits
    taking all water models into account. The data corresponding to the 
    lowest temperatures are found on right side of this diagram.}
  \label{fig:n13}
\end{figure}

\begin{figure}[!t]
  \centering
  \includegraphics[angle=0,width=7.4cm]{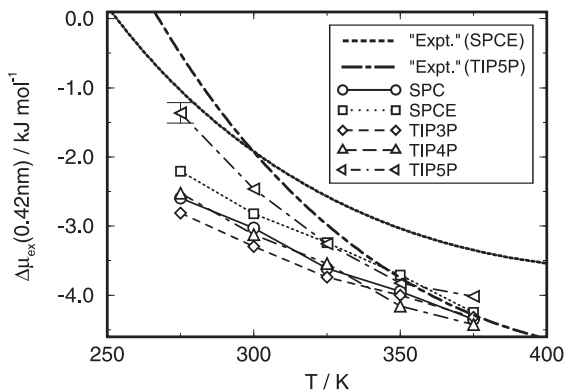}
  \caption{\footnotesize
    Excess chemical potential of Xenon 
    in the vicinity of another Xenon particle relative to the bulk.
    The distance corresponds to the
    maximum of the Xe-Xe $g(r)$-function ($r\!=\!0.42\;\mbox{nm}$).
    The heavy dashed and dot-dashed lines represents 
    the model prediction for the
    association of two
    Xenon particles in real water based on the MIT-parameters for
    SPCE and TIP5P (see text for details).}
  \label{fig:n14}
\end{figure}

\fi
In Figure \ref{fig:n12} we have applied the rescaling procedure proposed in
section \ref{sec:MIT}
to the
chemical potentials of the Xenon particles in contact with another Xenon particle.
Again we obtain an apparently linear behaviour for each of the water models, although
we must admit
that the corresponding slopes are found to vary more strongly than for
the case of the bulk liquid. 
However, it is quite evident that all slopes are
considerably smaller than the ones obtained for the bulk.
The fit parameters are given
in Table \ref{tab:MIT}.
An obvious conclusion is of course that the lines 
corresponding to bulk and shell
might have an intersection, defining the temperature where the interaction
between the hydrophobic particles turns from {\em attractive} into 
{\em repulsive}. 
The strongest evidence for this scenario comes from the TIP5P data
with a value of
$\Delta\mu_{ex}(0.42\,\mbox{nm})\!=\!-1.36\,\mbox{kJ}\,\mbox{mol}^{-1}$ 
at $275\,\mbox{K}$, which is actually
{\em weaker} than the pure Lennard Jones attraction between the two Xenon
particles, indicating  an already repulsive water cavity potential. 
In addition, the destabilisation of the 
contact-state with decreasing temperature is even enhanced by the increased
lowering of the solvent separated minimum of the profile of free energy
as shown in Figure \ref{fig:n09}.
The calculated intersection temperature for TIP5P water 
is found to lie quite high with $257\,\mbox{K}$. For the
other water models, however, the prediction of the intersection temperature is much
more problematic since the temperatures are apparently shifted to
even lower values. Hence the density fits, necessary to determine the
intersection temperatures, are much less precise.
Therefore the values of about $220\,\mbox{K}$ 
for SPC and TIP3P
and $230\,\mbox{K}$ for SPCE and TIP4P are subject to large uncertainty.

\ifpretty
\begin{figure}[!t]
  \centering
  \includegraphics[angle=0,width=7.0cm]{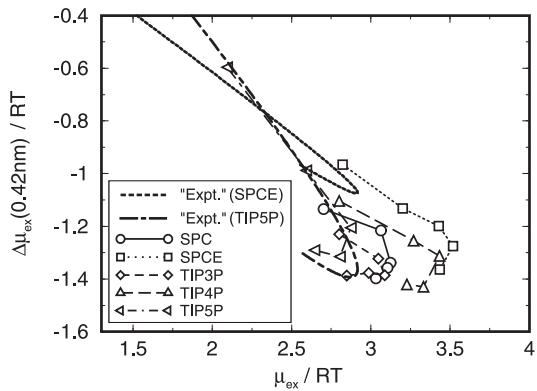}
  \caption{\footnotesize
    Variation of the strength of the hydrophobic interaction
    $\Delta \mu(0.42\mbox{nm})\!=\!\mu_{ex}(0.42\mbox{nm})-\mu_{ex}(\infty)$ with
    the hydration free energy $\mu_{ex}$. 
    The heavy dashed and dot-dashed lines represents 
    the model prediction for the
    association of two
    Xenon particles in real water (shown in Figure \ref{fig:n14}) based on the MIT-parameters for
    SPCE and TIP5P. }
  \label{fig:n15}
\end{figure}

\fi
In section \ref{sec:MIT} we have shown that the experimental data as well
as the TIP5P  and SPCE data can be almost quantitatively interrelated by the
MIT model assuming an effective Lennard Jones
sigma for the experimental water of 
$\sigma_{\rm Xe-Water}\!=\!3.5375\,\mbox{\AA}$. 
When applying the same procedure here, however, we should get an impression of how
the strength of the hydrophobic interaction of Xenon in real water might behave.
We would like to point out that the SPCE  and the TIP5P model represent the most extreme cases
with respect to their slopes shown in Figure \ref{fig:n13}.
The so calculated strengths of the hydrophobic interaction as a function of temperature are
shown in Figure \ref{fig:n14} and compared with the corresponding
data for the water models. Besides the large uncertainties of 
 this approach, the most striking feature,
the  strongly changing slope of the $\Delta \mu_{ex}$ curve, is mostly
due to the temperature dependence of waters expansivity. Hence
this observation does not depend on the exact values of
the $a$ and $c$ parameters, but requires just an approximately linear
dependence between $\mu_{ex}\,V_m^2$ and $V_m/T$, as it is suggested by
all five water models.
The corresponding temperatures for the attractive/repulsive
conversion are hence found to lie in
the interval between $253\,\mbox{K}$ and $267\,\mbox{K}$.

Recently Widom et al. \cite{Widom:2003} have deduced from lattice model
calculations the presence of an almost linear
relation between the hydrophobic interaction $\Delta \mu_{ex}/RT$ and the
free energy of hydrophobic hydration $\mu_{ex}/RT$ for
the limited temperature interval between $273\,\mbox{K}$ and $333\,\mbox{K}$.
Figure \ref{fig:n15} shows a such a plot containing the data obtained here for
the different water models. Figure \ref{fig:n15} indicates that at least for
temperatures sufficiently below the maximum of $\mu_{ex}(T)/RT$ their
observation is consistent with our data.
For the case of the SPCE, TIP4P and TIP5P models an almost
linear behaviour in the interval $275\mbox{K}$ and $325\mbox{K}$ is observed.
Please note that the model predictions for ``real Xenon in water'' based
 MIT model as outlined above behave as well
approximately linear for the lower temperatures, suggesting that the 
MIT predictions for the lower temperatures are as well conceptually consistent with the theory 
of Widom et al. \cite{Widom:2003}. We would like to point out that 
the slope of $0.7$ observed by Widom et al.
for Methane is closer to the value of $0.8$ obtained for the TIP5P model 
than to the $\approx0.4$ found for SPCE and the other water models.

%%% Local Variables: 
%%% mode: latex
%%% TeX-master: "paper"
%%% End: 

\section{CONCLUSIONS}

From a series of Molecular Dynamics simulations on 
different water models we conclude that the differences 
between the model waters structure, expansivity behavior and
the temperature dependence
of the solubility of simple solutes are tightly related. 
The calculated solvation entropies  
for all models are found to be systematically smaller than
the corresponding experimental values.
However, the water models (SPCE, TIP4P, TIP5P) that reproduce 
waters structure more accurately, provide solvation entropies
that are closer to the experimental data.

According to a modification of the Information theory model of Hummer et al.
we observe an almost linear dependence between $\mu_{ex}\,V_m^2$ and $V_m/T$
for both experimental and calculated excess chemical potentials.
A corresponding rescaling procedure is 
able to almost eliminate the
density dependencies of the different data. 
Moreover, when taking the size of Xenon particle according to the
different Lennard-Jones sigmas for the different models
into account, an almost quantitative prediction of the excess
chemical potential of Xenon for the different models and the experimental
data is feasible.

Concerning the hydrophobic interaction 
we would like to emphasise the simple fact that all
models show a qualitatively similar behaviour: Enhanced aggregation at elevated
temperatures. This is in general consistent with the simple picture of the
association being driven by entropy effects determined as
at net result of overlapping hydrations shells.
Nevertheless, from a  quantitative point of view noticeable differences 
between the different models exist,
which can only be rationalised considering molecular details of the hydration
shell in the Xenon-Xenon contact state.
However, in general we can
state that the water models that reveal a more realistic hydration entropy
and water structure, namely SPCE,
TIP4P and TIP5P, reveal as well a more strongly
pronounced association behaviour.
The TIP5P model reveals an extremely pronounced
temperature dependence.

The apparent linear behaviour between $\mu_{ex}\,V_m^2$ and $V_m/T$ is also 
found for Xenon in the Xenon-Xenon
contact state, strongly suggesting
the existence of a temperature where the hydrophobic interaction turns from
attractive into purely repulsive. 
For the case of TIP5P water this temperature is found to lie
quite high at about $257\,\mbox{K}$.
Moreover, the strong temperature
dependence of (experimental) waters expansivity strongly indicates
that the
weakening of the hydrophobic interactions in real water is
closer to that obtained for the TIP5P model.
Consequently, a model that accounts for waters density effects
as closest as possible is as well desirable for a correct 
description of hydrophobic interactions.

The almost linear relationship between hydrophobic interaction and the
strength of hydrophobic hydration proposed by
Widom et al. \cite{Widom:2003} is at least for temperatures
sufficiently below the maximum of $\mu_{ex}(T)/RT$ 
consistent with our data.

%%% Local Variables:
%%% mode: latex
%%% TeX-master: "paper"
%%% End:

\section*{ACKNOWLEGEMENT}

I am grateful to
Alfons Geiger and Ivan Brovchenko for helpful discussions.
Financing by the Deutsche Forschungsgemeinschaft
(DFG Forschergruppe 436) is gratefully acknowledged. 

\begin{appendix}

\section{Appendix: Check on the applicability of the particle insertion
  technique by  the overlapping distribution method
  in the isobaric-isothermal ensemble}
\label{sec:Appendix}

The Widom particle insertion method is 
known to fail in some cases, such 
as e.g. the high density Lennard-Jones liquid \cite{FrenkelSmit}.
In order to confirm that
the Widom method can be applied under the conditions of our
study we have used additional simulations 
of one Xenon particle in 500 water molecules
to calculate the chemical potential 
for Xenon using the overlapping distribution method 
as outlined in \cite{FrenkelSmit,LandauBinder}.
The additional 
simulations were conducted under the same temperatures/pressures
and simulation 
parameters as outlined in section \ref{sec:MD} and extended over
10 ns. Since Xenon is the  largest particle in our study, it has the
smallest free volume and 
therefore represents  the worst
case scenario comparing with other (smaller) noble gases and Methane.

\ifpretty \begin{figure}[!b]
  \centering
  \includegraphics[angle=0,width=6.8cm]{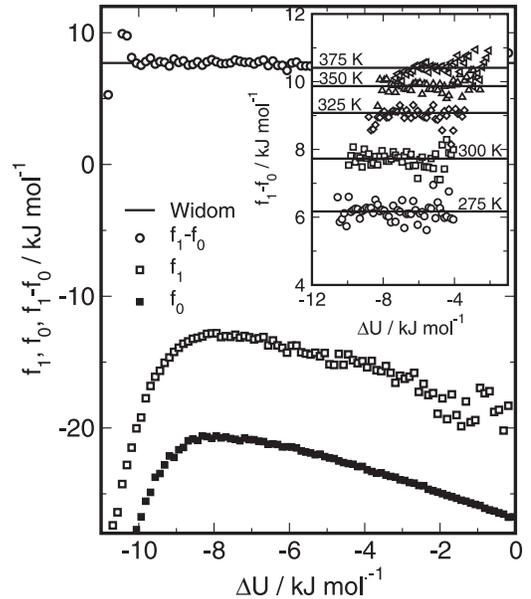}
  \caption{\footnotesize
    Excess chemical potential $\mu_{ex}$ according to
    the Widom particle insertion
    method and the method of overlapping distribution functions.
    All data were obtained
    at $p\!=\!0.1\;\mbox{MPa}$ under NPT-conditions. The figure body 
    shows
    $f_0$, $f_1$, $f_1-f_0$ functions and the Widom estimate
    at $T\!=\!300\;\mbox{K}$. 
    Insert: Well overlapping parts of $f_1-f_0$ and Widom estimates for
    all temperatures. All data were obtained from 
    simulations containing 500 SPCE
    molecules.}
  \label{fig:n16}
\end{figure}
 \fi
Since we are
dealing with simulations in the isobaric-isothermal ensemble,
the definition of the distribution functions
has to be slightly different than that for the canonical ensemble,
although this difference is usually ignored
\cite{Vlot:99}.
In analogy to the formulation of the overlapping distribution functions
in the canonical ensemble \cite{FrenkelSmit} 
we define two distribution functions (or
histograms) $p_1$ and $p_0$ of the energy of the $N+1$'th particle 
$\Delta U\!=\!U(\vec{s}^{N+1};L)-U(\vec{s}^{N};L)$. Here
$p_1(\Delta U)$ denotes the distribution of energies of the $N+1$'th
particle in the $N+1$-particle system in the isobar isothermal ensemble
\begin{eqnarray}
\label{eq:p1}
p_1(\Delta U)
& = &
\frac{1}{Q(N+1,P,T)}\;
\frac{\beta P}{\Lambda^{3(N+1)}\,N!} \;\times \nonumber\\
~&~ & 
\int dV
\int d\vec{s}_{N+1}
\int d\vec{s}^{N} \;V^{N+1} \times \nonumber\\
~&~ &
\exp(-\beta P V)\;
\exp\left[-\beta U(\vec{s}^{N+1};L)\right] \; \times \nonumber \\
~&~ &
\delta \left( U(\vec{s}^{N+1};L)-U(\vec{s}^{N};L) -\Delta U \right) 
\end{eqnarray}
where $\beta\!=\!1/kT$ and
$\vec{s}_N\!=\!L^{-1}\,\vec{r}_N$ (with $L\!=\!V^{1/3}$ being the length of 
a hypothetical cubic box) represent the scaled coordinates  of particle $N$. 
$\vec{s}^N$ and $\vec{s}^{N+1}$ represent the set of coordinates of the entire
$N$-and $N+1$-particle
systems, $P$ is the pressure and $V$ the volume. We have to denote that
the partition function contains the factor $N!$ instead of $(N+1)!$ here, since the
$N+1$'th particle, the solute particle, 
is distinguishable from the $N$
solvent particles in case it corresponds to a different particle type.
$U(\vec{s}^{N};L)$ and $U(\vec{s}^{N+1};L)$ denote the potential energies of the
$N$ and $N+1$ particle systems, respectively. $Q(\ldots)$ is the 
isobaric isothermal partition function.
Eq. \ref{eq:p1} is a straightforward extension of the expression for the 
canonical ensemble \cite{FrenkelSmit}.
The function $p_0(\Delta U)$ describes 
the energy-distribution of an additional
$N+1$-particle randomly inserted into 
configurations
representing an isobaric-isothermal ensemble of the
$N$-particle system.
In contrast to the $p_1(\Delta U)$ distribution
function, the
$p_0(\Delta U)$ distribution, however, 
is defined with the instantaneous volume being used as a
weighting factor
\begin{eqnarray}
p_0(\Delta U)
& = &
\frac{1}{Q(N,P,T)}\;
\frac{1}{\left<V\right>}\;
\frac{\beta P}{\Lambda^{3N}N!} \;\times\nonumber\\
~&~ &
\int dV
\int d\vec{s}_{N+1}
\int d\vec{s}^{N}\; V^{N}\;\times\nonumber\\
~&~ &
\exp(-\beta P V)
\exp\left[-\beta U(\vec{s}^{N};L)\right]\; \times\nonumber\\
~&~ &
V\;\delta \left( U(\vec{s}^{N+1};L)-U(\vec{s}^{N};L) -\Delta U \right)\;,
\end{eqnarray}
which has consequently to be normalised by the average volume $\left<V\right>$.
Starting with the definition of the $p_1(\Delta U)$-distribution function
and inserting the definition for $\Delta U$
we come up with the following relation between the two distribution functions
\begin{eqnarray}
p_1(\Delta U)
& = &
\frac{Q(N,P,T)}{Q(N+1,P,T)}\;
\frac{\left<V\right>}{\Lambda^{3}} \;\times \nonumber\\
& = &
\exp(-\beta \Delta U)\;
p_0(\Delta U) \;.
\end{eqnarray}
Using the defition of the {\em ideal} and {\em excess} part
of the chemical potential $\mu$ referring to
the ideal gas state with the same  average number density
in Eq. \ref{eq:muexdef}, we obtain
a relation between the two distribution functions and the excess chemical
potential which is analogous to the expression for the canonical ensemble
\begin{eqnarray}
\ln p_1(\Delta U) - \ln p_0(\Delta U) 
& = &
\beta \mu_{ex} -\beta \Delta U \;.
\end{eqnarray}
We have to emphasise  that the difference, however,
is the necessity of {\em volume-weighting}
in the calculation of the $p_0(\Delta U)$-distribution function.
As it is usually done \cite{FrenkelSmit}, 
we define functions $f_0$ and $f_1$ using
\begin{eqnarray}
f_0(\Delta U) 
& = & \ln p_0(\Delta U) - \frac{\beta \Delta U}{2} \;\;\;\hspace*{2em}\mbox{and}\nonumber\\
f_1(\Delta U)
& = & \ln p_1(\Delta U) + \frac{\beta \Delta U}{2} \nonumber
\end{eqnarray}
such that
\begin{eqnarray}
\beta \mu_{ex} & = &
f_1(\Delta U)-f_0(\Delta U)\;.
\end{eqnarray}
In most cases the relative volume-fluctuation will be small, and
therefore the volume-weighting will cause only a minor modification of to
the $p_0(\Delta U)$ distribution. However, when considering
states close to the critical point, where volume fluctuations 
might become large,
a significant influence cannot be ruled out. 
Nevertheless, the above formulation should be
used in any case, since the volume weighting can be done with practically no
additional computational effort.

In Figure \ref{fig:n16} the  $f_1$ and $f_0$ distribution functions obtained
from the $300\;\mbox{K}$ simulation are shown, as well as
difference between both functions and 
the excess chemical potential obtained from the particle insertion
method.
The $f_1$ and $f_0$ distribution functions 
overlap largely
and therefore a rather precise estimation of the excess chemical 
potential 
of about $\pm0.05\;\mbox{kJ}\,\mbox{mol}^{-1}$ 
is feasible. 
Figure \ref{fig:n16} (see insert) shows a detailed comparison of
$f_1-f_0$-data  values according to the Widom method
for all temperatures.
The values obtained from
the Widom particle insertion method agrees quantitatively
with the data obtained from
to the overlapping distribution method
(see also Table \ref{tab:muex} for the data)
The nice agreement between both methods,
can be explained by the form of the $f_0$-function. The $f_0$
exhibits a maximum, which indicates that the insertion procedure
also explores the low-energy edge of the energy distribution 
with
appropriate statistics. As a consequence
there is no bias in sampling the energy space
and the Widom method provides reliable results. 
Employing the method of Flyvbjerg and Petersen \cite{Flyvbjerg:89}
we estimate an error of $\pm0.1\;\mbox{kJ}\,\mbox{mol}^{-1}$ as
an upper bound for the accuracy of the Widom data for Xenon 
obtained from the 500 molecule system.

%%% Local Variables:
%%% mode: latex
%%% TeX-master: "paper"
%%% End:

\end{appendix}

%\bibliographystyle{apsrev}
%\bibliography{all}

\end{document}